\documentclass[aps,prx,twocolumn,showpacs,superscriptaddress,preprintnumbers,longbibliography]{revtex4-2} 
\usepackage{dcolumn}   % needed for some tables
\usepackage{bm}        % for math
\usepackage{amssymb}   % for math
\usepackage{amsmath}
\usepackage{comment}
\usepackage{color}
\usepackage{slashed}
\usepackage{braket}
\usepackage{amssymb,amsmath,color}
\usepackage[caption=false]{subfig}
\usepackage{graphicx}  % needed for figures
\usepackage[export]{adjustbox}
\usepackage[colorlinks=true,citecolor=blue,linkcolor=red]{hyperref}
\usepackage{multirow}
\usepackage{rotating,booktabs}
\usepackage[verbose]{placeins}
\usepackage{mathrsfs}
\usepackage{tikz}
\usetikzlibrary{quantikz}
\usepackage{stackengine}
\usepackage{soul}
\usepackage{float}

\newcommand{\Eq}[1]{Eq.~(\ref{#1})}

\newcommand{\Eqs}[2]{Eqs.~(\ref{#1}-\ref{#2})}
\newcommand{\Fig}[1]{Fig.~\ref{#1}}

\newcommand{\ua}{\mathord{\uparrow}}
\newcommand{\da}{\mathord{\downarrow}}

\makeatletter

\newcommand{\mycomment}[1]{}

\makeatletter
\def\blfootnote{\xdef\@thefnmark{}\@footnotetext}
\makeatother
\begin{document}

%\title{Measuring the Entanglement Structure of Non-Gaussian States}
\title{Entanglement Structure of Non-Gaussian States and How to Measure It}

\author{Henry Froland}
\email{frolandh@uw.edu}
\affiliation{InQubator for Quantum Simulation (IQuS), Department of Physics, University of Washington, Seattle, WA 98195, USA.}

\author{Torsten V. Zache}
\affiliation{Institute for Theoretical Physics, University of Innsbruck, Innsbruck, 6020, Austria}
\affiliation{Institute for Quantum Optics and Quantum Information of the Austrian Academy of Sciences, Innsbruck, 6020, Austria}

\author{Robert Ott}
\affiliation{Institute for Theoretical Physics, University of Innsbruck, Innsbruck, 6020, Austria}
\affiliation{Institute for Quantum Optics and Quantum Information of the Austrian Academy of Sciences, Innsbruck, 6020, Austria}

\author{Niklas Mueller}
\affiliation{InQubator for Quantum Simulation (IQuS), Department of Physics, University of Washington, Seattle, WA 98195, USA.}
\affiliation{Center for Quantum Information and Control, University of New Mexico, Albuquerque, NM 87106, USA}
\affiliation{Department of Physics and Astronomy, University of New Mexico, Albuquerque, NM 87106, USA}

\begin{abstract}
Rapidly growing capabilities of quantum simulators to probe quantum many-body phenomena  require new methods to characterize increasingly complex states. We present a protocol that constrains quantum states using experimentally measured correlation
functions. This method enables measurement of a quantum state's entanglement structure, opening a new route to study entanglement-related phenomena. Our approach extends Gaussian state parameterizations by systematically incorporating higher-order correlations. We show the protocol's usefulness in conjunction with current and forthcoming experimental capabilities, focusing on weakly interacting fermions as a proof of concept. Here, the lowest non-trivial expansion quantitatively predicts early time thermalization dynamics, including signaling the on-set of quantum chaos indicated by the entanglement Hamiltonian.
\end{abstract}
\maketitle

\noindent

\emph{Introduction.}--- 
Recent advances in quantum simulation have opened the door to understanding quantum many-body systems at unprecedented levels~\cite{gross2017quantum,browaeys2020many,altman2021quantum,beck2023quantum,bauer2023quantum,bauer2023quantuma}. These advances have come with a bevy of methods for characterizing quantum states from experimental measurements, ranging from full quantum state tomography~\cite{cramer2010efficient,flammia2012quantum}, over randomized measurement protocols~\cite{brydges2019probing,elben2023randomized,huang2020predicting}
and (entanglement) Hamiltonian learning~\cite{dalmonte2018quantum,bairey2019learning,anshu2021sample,kokail2021quantum,joshi2023exploring,
mueller2023quantum}, 
to many-body interference~\cite{daley2012measuring,pichler2013thermal,pichler2016measurement,kaufman2016quantum,islam2015measuring} or higher-order correlation functions~\cite{schweigler2017experimental,rispoli2019quantum}. 
An important goal of such methods is to reveal the structure of complex many-body states, including entanglement for characterizing quantum phases in equilibrium~\cite{li2008entanglement,dalmonte2022entanglement,bringewatt2024randomized} or chaos and thermalization~\cite{oganesyan2007localization,chang2019evolution,rakovszky2019signatures,mueller2022thermalization}.

Common approaches to measure entanglement structure incur costs in sample size or computational resources that grow exponentially with the (sub)system's size, which makes them challenging to scale up.
Moreover, these protocols are difficult to realize in experiments without full local or universal control, examples include analog quantum simulators based on ultra-cold atomic and molecular systems~\cite{bloch2008many,gross2017quantum,mazurenko2017cold,tarruell2018quantum}, or high-energy and nuclear physics experiments~\cite{beck2023quantum,bauer2023quantum,bauer2023quantuma,di2023quantum}. In such experiments, instead, one typically extracts information by measuring few-body correlation functions. 
\begin{figure}[t!]
    \centering
    \includegraphics[scale=.39, trim = 80 0 0 0]{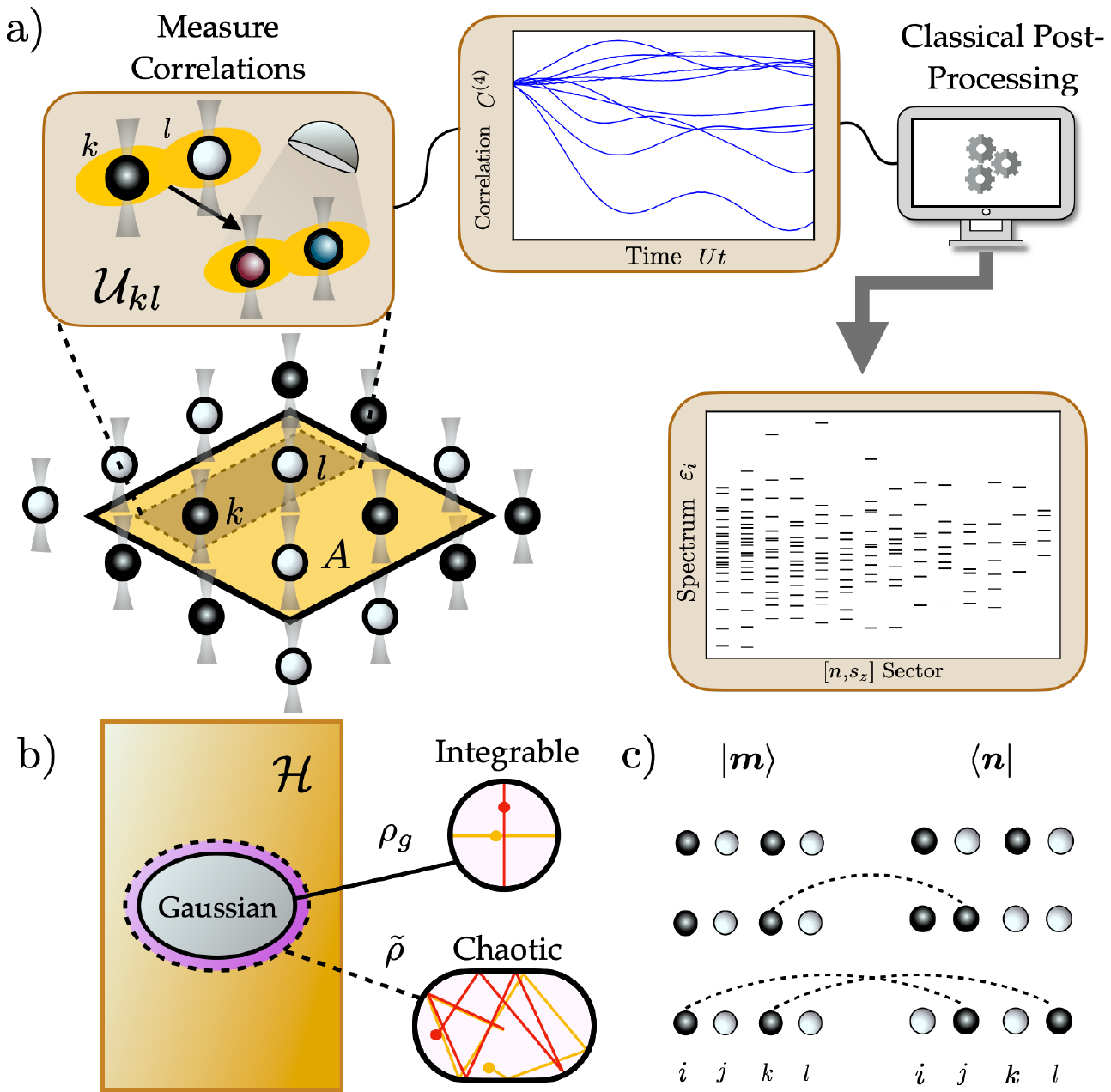}
        \caption{\textit{Non-Gaussian state reconstruction.} \textbf{(a)} Few-body correlation measurements (e.g. four-point correlations~$C^{(4)}$) on a subsystem $A$ of a fermionic quantum simulator are used to reconstruct the reduced state~$\rho_A$. Besides single-site resolved imaging, this requires pair-wise tunneling operations $\mathcal{U}_{kl}$ between sites $k$ and $l$. After classical post-processing, the protocol gives access to the entanglement spectrum $\varepsilon_i$. \textbf{(b)}~The state reconstruction is based on an expansion around Gaussian states $\rho_g$, which form a subset of integrable states in the Hilbert space~$\mathcal{H}$. To probe the onset of quantum chaos, we go to non-Gaussian states $\tilde{\rho}$ by including higher-order correlations. \textbf{(c)}~Four-point correlations lead to non-Gaussian matrix elements $\bra{\boldsymbol{m}}\rho_A\ket{\boldsymbol{n}}$ with Hamming distance $d_H(\boldsymbol{m},\boldsymbol{n})=0,2,4$ (top, middle, bottom).}
    \label{fig:fig1}
\end{figure}

Motivated by these restrictions, we address the following question in this letter: \textit{How much physics of entanglement can be captured if access is limited to lower order correlations}? Specifically, as illustrated in Fig.~\ref{fig:fig1}, we show how to utilize exact non-perturbative correlation functions accessible in experiments for constructing approximations to a quantum state based on an expansion of the many-body Wigner functional around Gaussian states. We expect our parameterization to capture accurate descriptions during early times, where higher correlations are small, and late times, particularly when the system is near thermal equilibrium. 

As an example, we illustrate our approach with numerically simulated thermalization dynamics of  a weakly-coupled extended Fermi-Hubbard chain after a quantum quench.
Here, we focus on the reconstruction of a non-equilibrium state's entanglement spectrum and analyze its level spacing distribution, which serves as a indicator of quantum chaos, a prerequisite for thermalization. Our work demonstrates how going beyond Gaussian state descriptions -- which are known to fail even qualitatively already for early times --
can quantitatively capture aspects of the chaotic evolution of the system's entanglement structure. Finally, we briefly discuss methods for probing such phenomena in analog fermion quantum simulators, based on existing and forthcoming experimental capabilities~\cite{gonzalez2023fermionic,impertro2023local,lunt2024realization,lebrat2024observation,prichard2024directly,hartke2023direct}.

\emph{Non-Gaussian States.}--- We discuss an approach for representing generic quantum states by expanding their Wigner functionals in terms of correlation functions. We consider a system of $N$ fermionic modes described by creation/annihilation operators $c^{\dagger}_i/c_i$ where $\{c^{\dagger}_i,c_j\}=\delta_{ij}$ ($i,j\in \{0,1,..,N-1\}$), and focus on connected two- and four-point functions,
\begin{subequations}
\begin{align}
    C^{(2)}_{ij} &\equiv \text{Tr}[ \rho\,  c_i^\dagger c_j]\\
    C^{(4)}_{ijkl} &\equiv \text{Tr} [\rho\,  c_i^\dagger c_j^\dagger c_k c_l]- C^{(2)}_{il} C^{(2)}_{jk}+C^{(2)}_{ik} C^{(2)}_{jl}\,.
\end{align}
\end{subequations}
For simplicity, we restrict ourselves to a U(1) symmetric state, e.g. a state with definite total particle number $N_F = \sum_i c^\dagger_i c_i$.
Correlations of this form are accessible in current and near-term neutral-atom experiments, illustrated in Fig. \ref{fig:fig1}(a) and discussed below~\footnote{It is not essential to deal with fermions, extending our approach to  bosonic models, and to higher order correlations, is straightforward.}. 

For the case of a subsystem defined by $N_A < N$ modes, the corresponding reduced state $\rho_A=\text{Tr}_{\bar{A}}\left(\rho\right)$ is equivalently described by a (bipartite) entanglement Hamiltonian (EH) $H_A \equiv -\log[\rho_A]$~\cite{li2008entanglement}, with entanglement spectrum~$\{\varepsilon_i\}$. For Gaussian states, the EH  is simply given by the covariance matrix, $C^{(2)}$, as $H_A =  -\sum_{ij} \log[C^{(2)} (1-C^{(2)})^{-1}]c_i^\dagger c_j+{\rm const}$~\cite{chung2001density}. This class of states falls short of capturing even the qualitative indicators of thermalization expected from the entanglement dynamics of the system, as we will argue below. Specifically, Gaussian states fall into a small ``corner" of the many-body Hilbert space $\mathcal{H}$ that cannot be chaotic, whereas even small deviations away from that region will show signatures of quantum chaos consistent with random matrix theory~\cite{guhr1998random,mehta2004random}, see \Fig{fig:fig1}(b).

Our approach is to write the state as a sum of a Gaussian  $\rho_g$, entirely determined by $C^{(2)}$, and non-Gaussian part $\delta \rho$,  
\begin{align}\label{eq:ansatz_def}
    {\rho} &\equiv \rho_g+\delta\rho
\end{align}
thereby reducing the problem to finding an expression for $\delta\rho$. To proceed, we utilize the formalism of fermionic coherent states \cite{cahill1999density} where $\rho$ may be expressed as
\begin{align}\label{eq:exact_state}
    \rho = \int {\rm d}^2\boldsymbol{\eta}{\rm d}^2\boldsymbol{\alpha}\;\;\
e^{W(\boldsymbol{\eta})} \;e^{\sum_{i}\alpha_i\eta_i^*-\eta_i\alpha_i^*}\ket{\boldsymbol{\alpha}}\bra{-\boldsymbol{\alpha}}
\end{align}
with $\boldsymbol{\alpha},\boldsymbol{\eta}$ being tuples of anti-commuting Grassmann numbers, $\boldsymbol{\alpha}\equiv \{\alpha_i,\alpha_i^* \dots\}_N$,  $\ket{\boldsymbol{\alpha}}$ are the corresponding coherent states, and $W(\boldsymbol{\eta})$ is the Wigner functional. We assume that $W(\boldsymbol{\eta})$ takes the form
\begin{align}\label{eq:gen_conn_corr}
    W(\boldsymbol{\eta}) = \sum_{i,j}C^{(2)}_{ij}\eta_j^{*}\eta_i+\frac{1}{4}\sum_{i,j,k,l}C^{(4)}_{ijkl}\eta^{*}_l\eta^{*}_k\eta_j\eta_i
\end{align}
with $i,j,k,l\in\{0,..,N_A-1\}$, which is justified for small $|C^{(4)}|\ll |C^{(2)}|$. We note that  $C^{(2)}$ and $C^{(4)}$ are not approximated and should be obtained from experiment, or computed  non-perturbatively~\footnote{Our ansatz is unrelated to time-dependent perturbation theory.}.
Working in the mode basis $\ket{\boldsymbol{n}}$ where $\rho_g$ is diagonal, (see Supplemental Material \cite{SM} for the full calculation where we also present a diagramatic approach) we derive explicit expressions for the matrix elements $\delta\rho_{\boldsymbol{m}\boldsymbol{n}}=\braket{\boldsymbol{m}|\delta\rho|\boldsymbol{n}}$,
\begin{widetext}
\begin{align}\label{eq:matrix_elts}
    \frac{\delta\rho_{\boldsymbol{m}\boldsymbol{n}}}{\rho_{g,\boldsymbol{n}\boldsymbol{n}}} =
    \begin{cases}
    \frac{1}{2}\sum_{i,j} (-1)^{n_i+n_j} \frac{\tilde{C}^{(4)}_{ijji}}{f_i^{\boldsymbol{n}}f_j^{\boldsymbol{n}}} &\text{if } d_H(\boldsymbol{m},\boldsymbol{n})=0\\
    \sum_{i,j,k} (-1)^{n_i + \varphi(j,k;\{i\},\boldsymbol{n})}
    \frac{\tilde{C}^{(4)}_{ijik}}{f_i^{\boldsymbol{n}}f_j^{\boldsymbol{n}}f_k^{\boldsymbol{n}}}
    \Gamma_{m_jn_k}^{n_jm_k} &\text{if } d_H(\boldsymbol{m},\boldsymbol{n})=2\\
    \frac{1}{4}\sum_{i,j,k,l} (-1)^{\varphi(i,j;\{k,l\},\boldsymbol{n})+\varphi(k,l;\{i,j\},\boldsymbol{n})}\text{sgn}(i-j)\text{sgn}(k-l)
    \frac{\tilde{C}^{(4)}_{ijkl}}{f_i^{\boldsymbol{n}}f_j^{\boldsymbol{n}}f_k^{\boldsymbol{n}}f_l^{\boldsymbol{n}}}
    \Gamma_{n_in_jm_km_l}^{m_im_jn_kn_l} &\text{if } d_H(\boldsymbol{m},\boldsymbol{n})=4
  \end{cases}
\end{align}
\end{widetext}
where $\rho_{g,\boldsymbol{n}\boldsymbol{n}}=\prod_pf_p^{\boldsymbol{n}}$ are the diagonal matrix elements of the Gaussian state and $f_p^{\boldsymbol{n}}\equiv n_pg_p+(1-n_p)(1-g_p)$, $g_p$ is the $p^{\mathrm{th}}$ eigenvalue of $C^{(2)}$, $\tilde{C}^{(4)}$ is the four-point correlation in the basis where $C^{(2)}$ is diagonal, and $d_H(\boldsymbol{m},\boldsymbol{n})$ is the hamming distance between the $\boldsymbol{m}$ and $\boldsymbol{n}$ occupation bitstrings. The symbol $\Gamma^{i_1,i_2,...,i_k}_{i'_1,i'_2,...,i'_{k'}}\equiv \prod_{a}^k\delta_{i_{a},0}\prod_{a}^{k'}\delta_{i'_{a},1}$ is defined so as to highlight the matrix structure of the $\delta\rho$ and is pictorially represented in Fig. \ref{fig:fig1}(c). The phases $\varphi(i,j;\{k,l\},\boldsymbol{n})\equiv \sum^{\max(i,j)-1}_{s=\min(i,j)+1}n_s$ count the occupation between modes $i$ and $j$  excluding the  modes $k$ and $l$. The parameterization exactly reproduces all correlations up to 1-body (2-point) and 2-body (4-point) and will be hereafter referred to as $\tilde{\rho}$.  We demonstrate 
its application using an example of a fermionic system.

\emph{Quench Setup.}-- We apply the non-Gaussian state parameterization to study the non-equilibrium dynamics after a quantum quench of a Fermi-Hubbard chain of spinful fermions with Hamiltonian~\cite{biebl2017thermalization}
\begin{align}
    H(J,J',U) &= J\sum_{l,s}c^{\dagger}_{l+1s}c_{ls} + J'\sum_{l,s}c^{\dagger}_{l+2s}c_{ls} + {\rm h.c.}\nonumber\\
    &\qquad+ U\sum_ln_{l\ua}n_{l\da}\label{eq:fermi_hubbard}
\end{align}
where $l$ labels the lattice site, $s=\ua,\da$ labels the spin, and $n_{ls}=c^{\dagger}_{ls}c_{ls}$ is the number operator~\footnote{Extended by next-to-nearest neighbor tunneling, this model is not integrable even in one spatial dimension}. Utilizing exact diagonalization~\cite{WeinbergED}, we initialize a random eigenstate of $H_0=H(1,1/8,0)$, which is Gaussian, and then evolve with $H(1,1/8,U)$ for a fixed value of interaction strength~$U$~\footnote{Different initial states are addressed in the Supplemental Material~\cite{SM}}. This quench setup allows us to study the build-up and evolution of non-Gaussian correlations. We monitor the entanglement structure of the exact time evolved state and the parameterization~$\tilde{\rho}$, utilizing exact correlation functions.

\emph{Early-time Dynamics.}---
In the early-time stage of the evolution, there are two regimes distinguished by the dynamics,  $\tau/N_A \ll 1$ and $\tau/N_A\ge \mathcal{O}(1)$ with $\tau = t U$. In the first regime, the four-point correlations grow quickly, with higher-order correlations being suppressed by powers of the interaction strength. For the second regime, higher-order correlations give a non-negligible contribution to $\rho_A$.
\begin{figure}
    \centering
    \includegraphics[scale=.43]{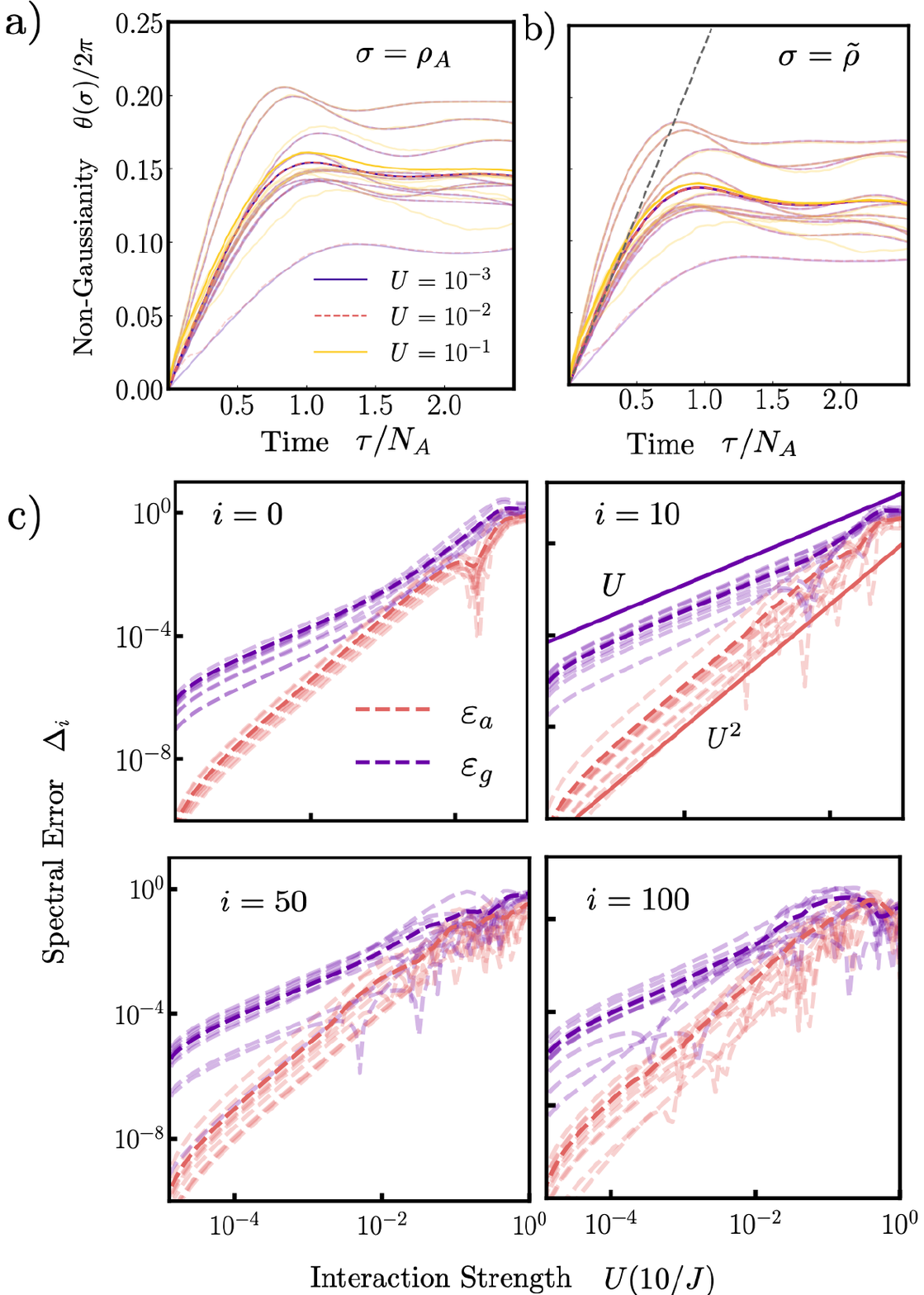}
    \caption{\textit{Non-equilibrium evolution of reconstructed state.} \textbf{(a,b)}~Non-Gaussianities~$\theta(\sigma)$, see Eq.~\eqref{eq:non-Gauss}, build up analogously for the exact state $\rho_A$ [(a)] and the reconstructed state~$\tilde{\rho}$~[(b)], as indicated by the dashed line which shows a linear fit to the early time behavior~$\theta(\rho_A)$. Here we show numerical data for a system of size $N=10,N_A=5$ with $10$ different initial states (shaded lines) and their averages (solid line). \textbf{(c)}~We test the accuracy of the state reconstruction through the spectral error of levels~$i$ of the entanglement spectra $\{\tilde{\epsilon}\}$, $\{\epsilon_g\}$ for different interaction strengths $U$ (at a fixed time). The non-Gaussian parameterization reduces the error to $\mathcal{O}(U^2)$ relative to the Gaussian parameterization with error $\mathcal{O}(U)$.}
    \label{fig:fig2}
\end{figure}
\begin{figure*}
    \centering
\includegraphics[scale=.6,trim=200 180 200 200]{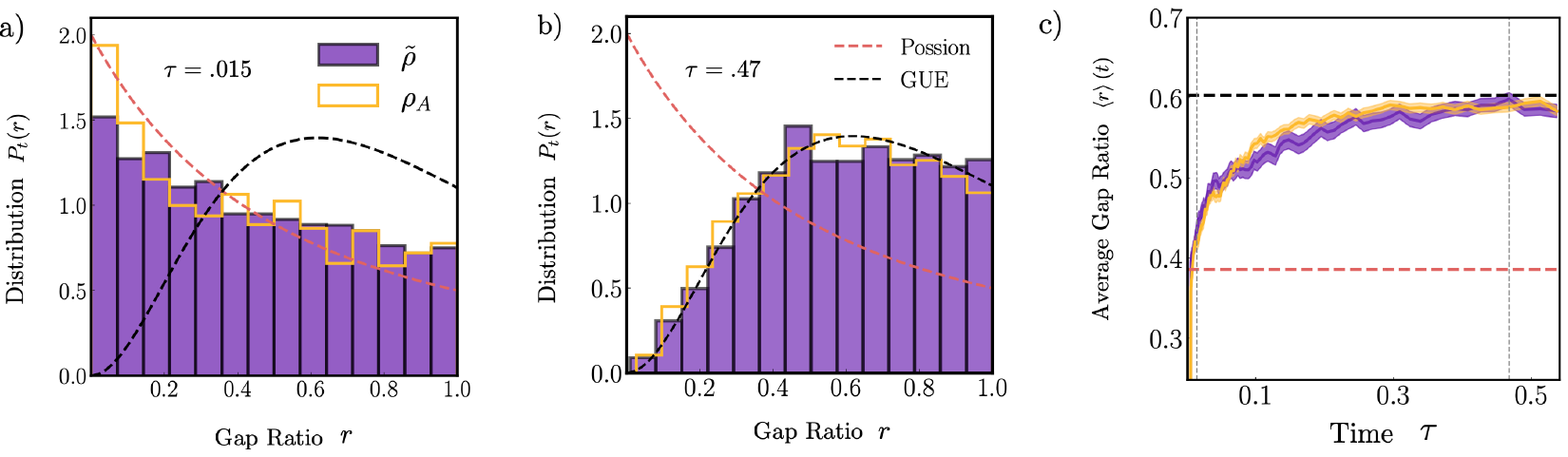}
    \caption{\textit{Evolution of level statistics of non-Gaussian state parameterization} \textbf{(a,b)}~ The gap ratio distributions of both, the parameterization $\tilde{\rho}$ (filled, purple) and the exact state $\rho_A$ (unfilled, gold), evolve from Poisson (pink dashed) at time $\tau=0.015$ to GUE (black dashed) at time $\tau=0.47$, thus signalling the onset of quantum ergodic evolution. \textbf{(c)}~The average gap ratio $\langle r \rangle$ grows with time and saturates to the value $\sim0.6$ indicating level repulsion (GUE). Data is shown for a system of size $N=10$, $N_A=5$, interaction strength $U=5.6\times10^{-3}$; averages are taken over 10 different initial states, where shaded regions signify $95\%$ confidence intervals. The dashed grey lines signify the times displayed in (a) and (b).}
    \label{fig:fig3}
\end{figure*}
\begin{figure}
    \centering
    \includegraphics[width=0.85\columnwidth]{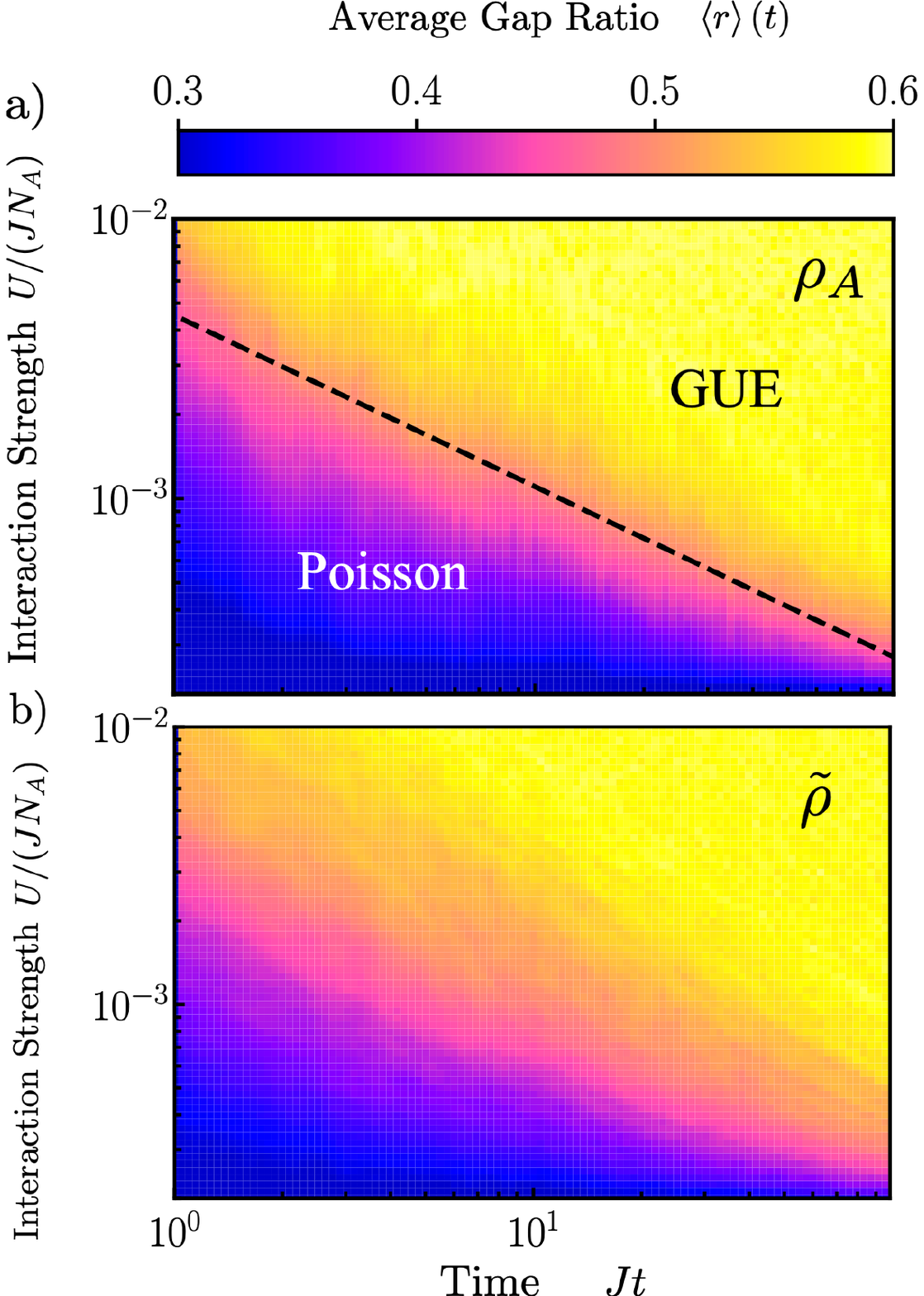}
    \caption{\textit{Average gap ratio of exact state and non-Gaussian parameterization} \textbf{(a)}~The average gap ratio of the exact state over a range of times $Jt$ and interaction strengths $U/J$. The black dashed line shows the crossover from a Poisson distribution to GUE, indicating quantum chaos through level repulsion. \textbf{(b)}~The average gap ratio for the parameterization qualitatively reproduces that of the exact state over the times and interaction strengths studied.}
    \label{fig:fig4}
\end{figure}%
We introduce a metric $\theta$ that measures the non-Gaussianity of a state,
\begin{align}
\label{eq:non-Gauss}
    \theta(\sigma)\equiv \arccos{\sqrt{\mathcal{F}(\sigma|\sigma_g)}}\,
\end{align}
where $\mathcal{F}(\sigma|\sigma') = \frac{\text{Tr}[\sigma\sigma']}{\text{max}\{\text{Tr}[\sigma^2],\text{Tr}[\sigma'^2]\}}$ is the max-fidelity~\cite{jozsa1994fidelity}, and $\sigma_g$ is the Gaussian part of $\sigma$ as in \Eq{eq:ansatz_def}. In \Fig{fig:fig2}, we show the deviation of the exact state from a Gaussian state, $\theta(\rho_A)$, in (a) and the deviation of the parameterization $\theta(\tilde{\rho})$ in (b), for both time regimes. At first the infidelity grows quadratically in $\tau$, i.e. $1-\mathcal{F}=\mathcal{O}(\tau^2)$, implying that both $\theta(\rho_A)$ and $\theta(\tilde{\rho})$ will grow linearly during this stage. The dashed line in (b) represents the linear part of $\theta(\rho_A)$ overlayed on $\theta(\tilde{\rho})$, highlighting that the growth is the same. Once the second stage of evolution has been reached, the expansion restricted to 1- and 2-body correlations is no longer quantitative, indicating that higher-order correlations not included in $\tilde{\rho}$ are now contributing to the dynamics. For small couplings $U\ll J,J'$ we also observe that $\theta$ has a self-similar profile in time (solid purple, dashed pink), which is inherited from the underlying correlations of $\rho_A$ (see SM for more details~\cite{SM}).

As a measure for how accurately the parameterization captures the entanglement spectrum $\{\epsilon_i\}$ of the exact state, we consider the spectral error 
\begin{align}
    \Delta_{i}=|\varepsilon_i-\varepsilon_{i}'|
\end{align}
where $\{\varepsilon_i'\}$ is the entanglement spectrum of either~$\tilde{\rho}$ or~$\rho_g$. In \Fig{fig:fig2}(c), we show the spectral error between~$\rho_g, \tilde{\rho}$ and~$\rho_A$ in levels $i=0,10,50,100$ as a function of $\tau$ with $t=10/J$ fixed. The figure shows that, for small interaction strength, the spectral error between $\rho_g$ and $\rho_A$ (dashed, pink) behaves as $\mathcal{O}(U)$, while the spectral error between  $\tilde{\rho}$ and $\rho_A$ (dashed, purple) scales as $\mathcal{O}(U^2)$. Thus $\tilde{\rho}$ captures the leading order in $U$ dependence of $\rho_A$, and  furnishes a better approximation of the exact state than just $\rho_g$. For larger interaction strengths neither $\rho_g$ nor $\tilde{\rho}$ accurately estimate the spectrum of $\rho_A$.

\emph{Thermalization Dynamics.}---
The statistical distribution of  the entanglement spectrum $\{ \varepsilon_i\}$ is an indicator of quantum ergodicity and chaos~\cite{guhr1998random}.
Specifically, we consider the gap ratio \cite{oganesyan2007localization}
\begin{align}
    r_i = \frac{\text{min}(\delta_i,\delta_{i-1})}{\text{max}(\delta_i,\delta_{i-1})}
\end{align}
where $\delta_i=\varepsilon_{i+1}-\varepsilon_i$. Figs.~\ref{fig:fig3}(a),(b) show the gap ratio distributions $P_t(r)$ of $\rho_A$ and $\tilde{\rho}$ averaged across symmetry sectors and different initial states~(see~\cite{SM} for computational details). These are compared to both a non-repulsive Poisson distribution and a Gaussian Unitary Ensemble (GUE). At early times the level statistics of both the exact state and parameterization are consistent with Poisson behavior while subsequent times show (GUE) level repulsion. Fig.~\ref{fig:fig3}(c) shows the average gap ratio $\braket{r}(t)$ of the parameterization versus the exact state over time, demonstrating that the level statistics of the exact state are reproduced. Fig.~\ref{fig:fig4} shows the average gap ratio over a range of interaction strengths~$U/J$ and time~$Jt$, comparing  both $\rho_A$ in a) with $\tilde{\rho}$ in b). Even at large $U$ the parameterization captures the level statistics of the exact state, and  accurately reproduces the  transition line between Poisson and GUE distributions. 

A Gaussian state cannot exhibit level repulsion because its EH is quadratic and parameterized solely by~$C^{(2)}$. Regardless of $C^{(2)}$, the EH cannot show level repulsion, except in the single-particle sector, simply because of the inherent level degeneracies in the many-particle sectors of a quadratic Hamiltonian. The incorporation of small $C^{(4)}$s is sufficient to break this degeneracy after short time evolution. The time scale at which this happens, nearly identical for the exact and the parametrized state, reveals a crucial insight: the emergence of quantum chaos necessitates the presence of two-body correlations of a certain magnitude.

\emph{Experimental Application.}---
Our approach can be applied in conjunction with current ultracold-atom simulators to address entanglement structure in experiments. We consider platforms which load fermionic atoms into optical lattices to realize Hubbard models with tunable nearest-neighbor hopping and on-site interaction strengths $U$~\cite{bloch2012quantum,mazurenko2017cold,leonard2023realization,gross2017quantum} similar to the model considered in Eq.~\eqref{eq:fermi_hubbard}.

To extract the entanglement structure using the above protocol, we require spin-resolved measurements of both diagonal fermion density correlations $\langle n_i n_j \rangle$ as well as off-diagonal fermion correlations $\langle c^{\dagger}_ic_j\rangle$, $\langle c^{\dagger}_ic^{\dagger}_jc_kc_l\rangle$ across the subsystem, see~\Fig{fig:fig1}(a). Besides site-resolved imaging techniques, such as quantum gas microscopes~\cite{browaeys2020many}, this requires the ability of realizing tunneling operations between arbitrary pairs of sites $i$ and $j$ in the subsystem. The required tunneling operations can for instance be efficiently implemented using optical tweezers~\cite{gonzalez2023fermionic,bluvstein2022quantum}.

To understand how one can use pair-wise tunneling to measure off-diagonal correlations, we write the operator ${c^{\dagger}_ic_j}={S^{x}_{ij}}+i{S^{y}_{ij}}$, where $S^{x}_{ij}=\frac{1}{2}(c^{\dagger}_ic_j+c^{\dagger}_jc_i)$ and $ S^{y}_{ij}=\frac{-i}{2}(c^{\dagger}_ic_j-c^{\dagger}_jc_i)$. By applying $\pi/2$-tunneling operations, $\mathcal{U}^{x}_{ij}(\pi/2)=\exp\{-i\frac{\pi}{4} S^{x}_{ij}\}$ and $\mathcal{U}^{y}_{ij}(-\pi/2)=\exp\{i\frac{\pi}{4} S^{y}_{ij}\}$, we bring $S^x_{ij}$ and $S^y_{ij}$, and thus ${c^{\dagger}_ic_j}$, into diagonal form where they are read from measurements of the local occupation numbers. The same technique can be applied to another set of sites in parallel to obtain the four-body operators. In general, one needs to perform $N_A(N_A-1)$ measurements for one-body correlations, where $N_A$ is the number of modes in the subsystem (including spin), and $N_A(N_A-1)(N_A-2)(N_A-3) + O(N_A^2)$ measurements for two-body correlations. $k$-body correlations require measurements in $O(N_A^{2k})$ bases where for $k\rightarrow \infty$ one performs, at exponential cost, full tomography.
Furthermore, generalized measurement protocols based on random atomic beam splitter operations in optical lattices~\cite{naldesi2023fermionic,gluza2021recovering,zhao2021fermionic,low2022classical,denzler2023learning,tran2023measuring} can be applied in the present context.

The advantage of our method is the use of polynomial (in system size) number of correlation functions—standard observables in experiments—as input, rather than requiring full state tomography. However, depending on the specific application, a complexity analysis must also account for the sampling cost associated with estimating correlation functions to the desired precision and, if targeting the EH's level spacing distribution, the classical representation of the reduced density matrix, which we address in~\cite{SM}. Our analysis suggests that the spectral properties of the EH we target may be costly to resolve, potentially exponentially so—especially at early times, where higher-order correlations tend to be overestimated with finite shots. This direction has only recently begun to be addressed~\cite{gu2024simulating,bittel2024optimal}.

The accuracy of approximation is  dependent on the strength of correlations that the state exhibits, which one has no a priori knowledge of in a quantum simulation experiment. Convergence can be explicitly tested by measuring higher-order correlations and incorporating them into the parameterization, as described in~\cite{SM}.

\emph{Conclusions.}--- In this Letter we addressed the entanglement structure of quantum states, i.e. entanglement Hamiltonians and their spectrum, from the perspective of lower-order $k$-body correlation functions. Our approach is based on a systematic expansion of the Wigner function around Gaussian states, where non-Gaussian contributions are constructed directly from higher-order correlations. While our primary example focused on $k=$ 1 and 2, the method can be systematically extended to all orders. Furthermore, the approach is model-independent and applicable in higher dimensions. 

Our motivation for this study was multi-fold. On one hand, our approach provides a practical avenue for measuring the entanglement structure in analog quantum simulator experiments with limited control. Especially at weak-coupling, these experiments could enable the exploration of quantum chaos and thermalization at late times, connecting entanglement dynamics to the growth of lower-order correlations. We posit that the parameterization we have developed is useful at late times near-thermal equilibrium, where states are expected to be close to Gaussian. Thus our approach offers a pathway from the early to the late stages of the thermalization dynamics~\cite{deutsch1991quantum,srednicki1994chaos,rigol2008thermalization,yao20232,ebner2024eigenstate}. 

% From a different perspective, one area that remains unexplored is the connection between correlation functions and entanglement in high-energy and nuclear physics experiments. Specifically, in ultra-relativistic heavy ion collisions many-particle correlations are a key experimental observable~\cite{vogt2007ultrarelativistic,florkowski2010phenomenology}. Despite the lack of direct access to any off-diagonal correlations in experiment (relative to the particle basis), correlation functions play a central role in various theoretical descriptions, including effective kinetic theory~\cite{arnold2003effective} and hydrodynamics~\cite{shen2020recent}. Within these frameworks, that are validated and constrained by experimental data, our approach, generalized to bosonic content, may see use for studying physics encoded in the entanglement spectrum. Furthermore, the intersections of our approach with other theoretical tools available for studying quantum-many body dynamics, such as matrix product states~\cite{schollwock2011density}  or time-dependent variational techniques~\cite{kramer1981geometry,haegeman2011time} ought to be explored. Finally, non-Gaussianity can be viewed as a quantum information theoretic resource~\cite{zhuang2018resource,takagi2018convex,albarelli2018resource}. 
% Its role for measuring complexity of quantum many-body systems should be explored~\cite{hebenstreit2020computational,leone2022stabilizer,tirrito2024quantifying,rattacaso2023stabilizer,robin2024magic,hahn2024bridging,gu2024magic,mele2024learning}

One area that remains unexplored is the connection between correlation functions and entanglement in high-energy and nuclear physics experiments. In ultra-relativistic heavy ion collisions many-particle correlations are a key experimental observable~\cite{vogt2007ultrarelativistic,florkowski2010phenomenology}. Despite the lack of direct access to any off-diagonal correlations in experiment, correlation functions are central for theoretical descriptions, including effective kinetic theory~\cite{arnold2003effective} and hydrodynamics~\cite{shen2020recent}. Within these frameworks, our approach may see use for studying physics encoded in the entanglement spectrum. Furthermore, the intersections of our approach with other theoretical tools available for studying quantum-many body dynamics, such as matrix product states~\cite{schollwock2011density}  or time-dependent variational techniques~\cite{kramer1981geometry,haegeman2011time} ought to be explored. Our technique is more broadly applicable for characterizing quantum states without resorting to full tomography, e.g. to probe phases of matter through the low-energy spectrum of the EH~\cite{li2008entanglement} or to evaluate measures of quantum complexity~\cite{tirrito2024quantifying}. Finally, non-Gaussianity can be viewed as a quantum information theoretic resource~\cite{zhuang2018resource,takagi2018convex,albarelli2018resource}. 
Its role for measuring complexity of quantum many-body systems should be explored~\cite{hebenstreit2020computational,leone2022stabilizer,rattacaso2023stabilizer,robin2024magic,hahn2024bridging,gu2024magic,mele2024learning}.

\emph{Acknowledgements.}--
We thank Mark Rudner, Martin Savage, and Rahul Trivedi for discussions. We also thank the participants of the InQubator for Quantum Simulation (IQuS) workshop ``Thermalization, from Cold Atoms to Hot Quantum Chromodynamics'' (\url{https://iqus.uw.edu/events/iqus-workshop-thermalization/}) at the University of Washington in September 2023 for many valuable discussions leading to this work. 
H.F. and N.M. acknowledge funding by the DOE, Office of Science, Office of Nuclear Physics, IQuS (\url{https://iqus.uw.edu}), via the program on Quantum Horizons: QIS Research and Innovation for Nuclear Science under Award DE-SC0020970.
This work is supported by the European Union’s Horizon Europe research and innovation program under Grant Agreement No. 101113690 (PASQuanS2.1), the ERC Starting grant QARA (Grant No.~101041435), the EU-QUANTERA project TNiSQ (N-6001), and by the Austrian Science Fund (FWF): COE 1 and quantA. This work was enabled, in part, by the use of advanced computational, storage and networking infrastructure provided by the Hyak supercomputer system at the University of Washington \cite{uw_hyak}

\bibliography{refs}

\section*{Supplemental Material}

\section{Derivation of Parameterization}\label{app:derivation}
In this section of the Supplemental Material, we provide a detailed derivation of the parameterization used in this manuscript, filling in the steps between Eq.~(3) and Eq.~(5) in the main text, and illustrating a general approach that can be extended to include higher-order correlation functions. First, we will present some basic identities and clarify the conventions used. Then, we will derive Eq.~(5). Additionally, we introduce a diagrammatic formulation that aids intuition and can be straightforwardly extended beyond the two-body correlations demonstrated here. The  fermionic coherent state formalism for the Wigner functional follows the conventions of Cahill and Glauber~\cite{cahill1999density}. Our derivation is built on fermionic states  consisting of modes $i = 0,\dots, N-1$, with creation and annihilation operators obeying the usual anticommutation relation $\{ c_i, c_j^\dagger \} = \delta_{ij}$. Coherent states, denoted as $| \boldsymbol{\gamma} \rangle \equiv D(\boldsymbol{\gamma} ) |0 \rangle$, are defined by the action of the displacement operator $D(\boldsymbol{\gamma}) \equiv \exp( \sum_i c^\dagger_i \gamma_i - \gamma^*_i c_i )$. Here, $c_n |0 \rangle = 0$ defines the Fock vacuum, and $\boldsymbol{\gamma}\equiv (\gamma_0 ,\gamma_0^*,\dots)$ represents a tuple of $2N$ Grassmann variables that anticommute. Grassmann integration uses the convention that $\int d^2 \boldsymbol{\gamma}\equiv \int \prod_i d\gamma_i d\gamma_i^*$, and $\int d\gamma_i \gamma_j = \int d\gamma_i^* \gamma_j^*=\delta_{ij}$. We will make use of the completeness relation $\mathbb{I} = \int d^2{\gamma} \, \gamma\gamma^* \,D(\gamma) $ and other standard relations found in~\cite{cahill1999density}. Products of fermionic or Grassmann variables will be contain an arrow $ {\leftharpoonup}$ or $ {\rightharpoonup}{}$ to denote decreasing or increasing mode ordering, unless the ordering does not matter (such as for products of bi-linear terms). Derivatives, including the product rule, for Grassmann variables follow~\cite{cahill1999density}.

\subsection{Evaluation of non-Gaussian Wigner functional}
\noindent
Inserting Eq.~(4) of the main text into Eq.~(3) yields
\begin{multline}
    \rho_A = \int d^2\boldsymbol{\eta}d^2\boldsymbol{\alpha}\;\;\
    e^{\sum_{ij}C^{(2)}_{ij}\eta_j^*\eta_i}\\
    \left(1+ \frac{1}{4}\sum_{i,j,k,l}C^{(4)}_{ijkl}\eta^{*}_l\eta^{*}_k\eta_j\eta_i\right)
    \;e^{\sum_i\alpha_i\eta_i^*-\eta_i\alpha_i^*}\ket{\boldsymbol{\alpha}}\bra{-\boldsymbol{\alpha}}\\
    + \mathcal{O}\left(C^{(6)}\right)
\end{multline}
where only the leading order dependency in $C^{(4)}$ has been kept. This expression can be separated into two terms, $\rho_A \equiv \rho_g + \delta\rho$
\begin{align}
    \rho_g &= \int d^2\boldsymbol{\eta}d^2\boldsymbol{\alpha}\;\;\
e^{\sum_{ij}C^{(2)}_{ij}\eta_j^*\eta_i}e^{\sum_i\alpha_i\eta_i^*-\eta_i\alpha_i^*}\ket{\boldsymbol{\alpha}}\bra{-\boldsymbol{\alpha}}\,,\\
    \delta\rho &= \frac{1}{4}\sum_{i,j,k,l}C^{(4)}_{ijkl}\int d^2\boldsymbol{\eta}d^2\boldsymbol{\alpha}\;\eta^{*}_l\eta^{*}_k\eta_j\eta_ie^{\sum_{ij}C^{(2)}_{ij}\eta_j^*\eta_i}\nonumber\\
    & e^{\sum_i\alpha_i\eta_i^*-\eta_i\alpha_i^*}\ket{\boldsymbol{\alpha}}\bra{-\boldsymbol{\alpha}}\label{eq:pert_integral}
\end{align}
where the first line is identified with the Gaussian part of $\rho_A$~\footnote{Note that one can bring the Gaussian part into an alternative form~\cite{chung2001density}, noting that in the diagonal basis $\rho_g = \det(C^{(2)}) \exp\{-\sum_{ij} \log[C^{(2)}(1-C^{(2)})^{-1} c_i^\dagger c_j] \}  = \prod_i g_i [\frac{1-g_i}{g_i}]^{\hat{n}_i}$, where $\hat{n}_i$ are the number operators in the $C^{(2)}$ eigenbasis. Taking the matrix elements $\langle\boldsymbol{m} | \rho_g | \boldsymbol{n}\rangle$ yields the expression we work with in this manuscript.} and the second is the perturbation. We are interested in the matrix elements in the Fock basis, $\rho_{g,\boldsymbol{m}\boldsymbol{n}}\equiv \langle \boldsymbol{m} | \rho_g | \boldsymbol{n}\rangle$, $\delta\rho_{\boldsymbol{n}\boldsymbol{m}}\equiv\langle \boldsymbol{m} | \delta \rho | \boldsymbol{n}\rangle$, where we assume that $| \boldsymbol{n} \rangle \equiv \overset{\leftharpoonup}{\prod}_{{i}} (c_i^\dagger)^{n_i}\ket{0}
$ where $\boldsymbol{n}\equiv (n_0,\dots )$ with $n_i\in 0,1$, and the left-pointing arrow means that the modes are in descending order. To proceed, we change to the Grassmann coordinates $\xi_i=\sum_jU_{ij}\eta_j$, where $U$ is the matrix that diagonalizes $C^{(2)}$. With this change the integrals over $\boldsymbol{\xi}$ and $\boldsymbol{\alpha}$ factorize and the Gaussian part becomes
\begin{align}
    \rho_{g,\boldsymbol{m}\boldsymbol{n}} &= \int d^2\boldsymbol{\xi}d^2\boldsymbol{\alpha}\;
    e^{\sum_ig_i\xi_i^*\xi_i+\alpha_i\xi_i^*-\xi_i\alpha_i^*}\braket{\boldsymbol{m}|\boldsymbol{\alpha}}\braket{-\boldsymbol{\alpha}|\boldsymbol{n}}\nonumber\\
    &= \text{det}[-C^{(2)}]\int d^2\boldsymbol{\alpha} e^{\sum_i\frac{\alpha_i^*\alpha_i}{g_i}} \langle{\boldsymbol{m}|\boldsymbol{\alpha}}\rangle\langle{-\boldsymbol{\alpha}|\boldsymbol{n}}\rangle
\end{align}
where $\text{det}[-C^{(2)}]\equiv\prod_i (-g^{(2)}_i)$ and  $g_i$, $i=0,N-1$ are the eigenvalues of the correlation matrix $C^{(2)}$ and
\begin{align}
    \langle \boldsymbol{m} | \boldsymbol{\alpha} \rangle &= \overset{\rightharpoonup}{\prod_a} (\alpha_a)^{m_a}  \exp\{-\frac{\alpha_a^* \alpha_a}{2}\}\\
    \langle -\boldsymbol{\alpha} | \boldsymbol{n}\rangle &=\overset{\leftharpoonup}{\prod_a} \exp\{-\frac{\alpha_a^* \alpha_a}{2}\}   (-\alpha^*_a)^{n_a}\,.
\end{align}
The Gaussian matrix elements become
\begin{align}
    \rho_{g,\boldsymbol{m}\boldsymbol{n}} &=\text{det}[-C^{(2)}]\int d^2\boldsymbol{\alpha}\,e^{\sum_i[\frac{1}{g_i}-1]\alpha_i^*\alpha_i}\nonumber\\&\qquad\times \overset{\rightharpoonup}{\prod_a} (\alpha_a)^{m_a}\overset{\leftharpoonup}{\prod_a}(-\alpha^*_a)^{n_a}\nonumber\\
    &=\delta_{\boldsymbol{m}\boldsymbol{n}}\prod_i [n_ig_i+(1-n_i)(1-g_i)]\,,
\end{align}
where $\delta_{\boldsymbol{m}\boldsymbol{n}} \equiv \prod_i \delta_{{m_i}{n_i}}$. In the second equality we made use of the following identity,
\begin{align}
    \int d^2\alpha_i e^{\lambda\alpha_i^*\alpha_i}{\alpha_i}^{m_i}(-\alpha_i^*)^{n_i} &= \delta_{m_in_i}[\lambda-n_i(1+\lambda)]\label{eq:useful_int}\,.
\end{align}
After computing the Gaussian component, we proceed deriving the quartic contribution, $\delta\rho$ [\Eq{eq:pert_integral}], which incorporates two-body correlations. To simplify the derivation, we note first that a specific term in the integrand of \Eq{eq:pert_integral} can be written as
\begin{align}
    \eta^{*}_l \eta^{*}_k \eta_j \eta_i e^{\sum_i\alpha_i\eta_i^*-\eta_i\alpha_i^*} = \frac{\partial}{\partial \alpha_l} \frac{\partial}{\partial \alpha_k} \frac{\partial}{\partial \alpha^{*}_j} \frac{\partial}{\partial \alpha^{*}_i} e^{\sum_i\alpha_i\eta_i^*-\eta_i\alpha_i^*}
\end{align}
Integrating \Eq{eq:pert_integral} by parts in the $\boldsymbol{\alpha}$ coordinates moves the derivative terms to act on $\braket{\boldsymbol{m}|\boldsymbol{\alpha}}\braket{-\boldsymbol{\alpha}|\boldsymbol{n}}$. As with the Gaussian case, we switch to the $\boldsymbol{\xi}$ basis where $C^{(2)}$ is diagonal, denoting the four-point function in this basis as $\tilde{C}^{(4)}$ where $\tilde{C}^{(4)}_{ijkl}\equiv \sum_{abcd}U_{ia} U_{jb} {C}^{(4)}_{abcd}U^\dagger_{ck}U^\dagger_{dl}$. The $\boldsymbol{\xi}$ integrals may be evaluated as before, and  yet another integration by parts, this time in the $\boldsymbol{\alpha}$ variables, gives
\begin{multline}\label{eq:pert_elts}
    \delta\rho_{\boldsymbol{m}\boldsymbol{n}} = \frac{1}{4}\text{det}[-C^{(2)}]\sum_{ijkl}\tilde{C}^{(4)}_{ijkl} \int d^2\boldsymbol{\alpha}\\
    \left[\frac{\partial}{\partial \alpha_l} \frac{\partial}{\partial \alpha_k} \frac{\partial}{\partial \alpha^{*}_j} \frac{\partial}{\partial \alpha^{*}_i}\prod_pe^{\frac{\alpha_p^{*}\alpha_p}{g_p}}\right]\braket{\boldsymbol{m}|\boldsymbol{\alpha}}\braket{-\boldsymbol{\alpha}|\boldsymbol{n}}
\end{multline}
The derivative terms in Eq. (\ref{eq:pert_elts}) are evaluated using the product rule for Grassmann numbers,
\begin{align}\label{eq:derivatives}
    \frac{\partial}{\partial \alpha_l} \frac{\partial}{\partial \alpha_k} \frac{\partial}{\partial \alpha^{*}_j} \frac{\partial}{\partial \alpha^{*}_i}\prod_p &e^{\frac{\alpha_p^{*}\alpha_p}{g_p}} =
    \frac{\delta_{il}\delta_{jk}-\delta_{ik}\delta_{jl}}{g_ig_j}\prod_{p\neq i,j}e^{\frac{\alpha_p^{*}\alpha_p}{g_p}}\nonumber\\
    &+\frac{\delta_{jk}\alpha_i\alpha_l^*-\delta_{ik}\alpha_j\alpha_l^*}{g_ig_jg_l}\prod_{p\neq i,j,l}e^{\frac{\alpha_p^{*}\alpha_p}{g_p}}\nonumber\\
    &+\frac{\delta_{il}\alpha_j\alpha_k^*-\delta_{jl}\alpha_i\alpha_k^*}{g_ig_jg_k}\prod_{p\neq i,j,k}e^{\frac{\alpha_p^{*}\alpha_p}{g_p}}\nonumber\\
    &+\frac{\alpha_i\alpha_j\alpha_k^*\alpha_l^*}{g_ig_jg_kg_l}\prod_{p\neq i,j,k,l}e^{\frac{\alpha_p^{*}\alpha_p}{g_p}}
\end{align}
Here, in second line we implicitly assume that $i\neq j\neq l$, and in the third $i\neq j\neq k$, and in the fourth $i\neq j\neq k\neq l$, otherwise these terms vanish.
 Inserting \Eq{eq:derivatives} into \Eq{eq:pert_elts} gives
 \begin{align}\label{eq:Isum}
 \delta\rho_{\boldsymbol{m}\boldsymbol{n}}\equiv 2\sum_{i,j}I_1^{(i,j)}+4\sum_{i,j,k}I_{2}^{(i,j,k)}+\sum_{i,j,k,l}I_3^{(i,j,k,l)}\,,
 \end{align}
 where we abbreviated
 \begin{align}\label{eq:sm:i1}
    I_1^{(i,j)} = \nonumber\frac{1}{4g_ig_j}\text{det}[-C^{(2)}]\, \tilde{C}^{(4)}_{ijji} \int d^2\boldsymbol{\alpha}\prod_{p\neq i,j}e^{\frac{\alpha_p^{*}\alpha_p}{g_p}}\\
   \times \prod_qe^{-\alpha_q^*\alpha_q}\overset{\rightharpoonup}{\prod_a} (\alpha_a)^{m_a}\overset{\leftharpoonup}{\prod_b} (-\alpha_b^*)^{n_b}
\end{align}
When $\boldsymbol{m}\neq \boldsymbol{n}$, there will be an unpaired Grassmann number and the entire integral will vanish, consequently~\Eq{eq:sm:i1} is non-zero only if $\boldsymbol{m}=\boldsymbol{n}$. To compute the remaining integrals, one combines the Grassmann variables into bilinear form $\sim \alpha_i^* \alpha_i$ and, using \Eq{eq:useful_int}, one finds
\begin{align}\label{eq:I1_fin}
    I_1^{(i,j)}
    = \frac{1}{4}{\tilde{C}^{(4)}_{ijji}}&\delta_{\boldsymbol{m}\boldsymbol{n}}(-1)^{n_i+n_j}\nonumber\\&\times\prod_{p\neq i,j}[n_pg_p+(1-n_p)(1-g_p)]\,.
\end{align}

The second term of \Eq{eq:Isum} can be written as
\begin{align}
    I_2^{(i,j,k)} =\nonumber \frac{1}{4g_ig_jg_k}\text{det}[-C^{(2)}]\tilde{C}^{(4)}_{ijik}\int d^2\boldsymbol{\alpha}\;\alpha_j\alpha_k^*\\
    \times \prod_{p\neq i,j,k}e^{\frac{\alpha_p^{*}\alpha_p}{g_p}}\prod_qe^{-\alpha_q^*\alpha_q}\overset{\rightharpoonup}{\prod_a} (\alpha_a)^{m_a}\overset{\leftharpoonup}{\prod_b} (-\alpha_b^*)^{n_b}\label{eq:I2}\,.
\end{align}
The Grassmann integrals only evaluate to a non-zero result if they appear once and the integration is over bilinears of the same mode. This is only the case if $n_k=0$, $m_j=0$, $n_j=1$, and $m_k=1$, as well as $n_a=m_a$ for $a\neq i,j$.
To represent this condition we introduce the abbreviation $\Gamma_{m_jn_k}^{n_jm_k}$,
\begin{align}
\Gamma^{i_1,i_2,...,i_k}_{i'_1,i'_2,...,i'_{k'}} &\equiv \prod_{a}^k\delta_{i_{a},0}\prod_{a}^{k'}\delta_{i'_{a},1}\label{eq:delta_symbol}\,,
\end{align}
where an index placed in superscript indicates that the corresponding mode should be unoccupied $n_i=0$, and an index in subscript means that the mode is occupied $n_i=1$. A mode $i$ can only appear in either super- or subscript. We further use the following identity,
\begin{multline}
    \alpha_j\alpha_k^*\, \overset{\rightharpoonup}{\prod_a} (\alpha_a)^{m_a}\overset{\leftharpoonup}{\prod_b} (-\alpha_b^*)^{n_b} =\\
    (-1)^{\varphi(j,k;\boldsymbol{n})}\alpha_k^*\alpha_k\,\alpha_j^*\alpha_j\overset{\rightharpoonup}{\prod_{a\neq k}} (\alpha_a)^{m_a}\overset{\leftharpoonup}{\prod_{b\neq j}} (-\alpha_b^*)^{n_b}\label{eq:occ_phase}
\end{multline}
where $\varphi(j,k;\{i\},\boldsymbol{n})=\sum_{j<s<k}n_s$ if $j<k$ and $\varphi(j,k;\{i\},\boldsymbol{n})=\sum_{k<s<j}n_s$ if $j>k$, counting the occupation between modes $j$ and $k$, ignoring the occupation of mode $i$. Inserting this identity into \Eq{eq:I2} yields
\begin{align}
    I_2^{(i,j,k)} 
   &= \frac{1}{4}\tilde{C}^{(4)}_{ijik}(-1)^{\varphi(j,k;\{i\},\boldsymbol{n})} (-1)^{n_i}\Gamma_{n_jm_k}^{m_jn_k}\delta_{n_im_i}\nonumber\\
    &\times\prod_{p\neq i,j,k}\delta_{n_pm_p}[n_pg_p+(1-n_p)(1-g_p)]\label{eq:I2_fin}\,.
\end{align}
The final contribution is given by% $I_3^{(i,j,k,l)}$ is 
\begin{align}
    I_3^{(i,j,k,l)}
    = \frac{1}{4g_ig_jg_kg_l}\text{det}[-C^{(2)}]{\tilde{C}^{(4)}_{ijkl}}{}\int d^2\boldsymbol{\alpha}\;\alpha_i\alpha_j\alpha_k^*\alpha_l^*\nonumber\\\times\prod_{p\neq i,j,k,l}e^{\frac{\alpha_p^{*}\alpha_p}{g_p}}
    \prod_qe^{-\alpha_q^*\alpha_q}\overset{\rightharpoonup}{\prod_a} (\alpha_a)^{m_a}\overset{\leftharpoonup}{\prod_b} (-\alpha_b^*)^{n_b}\label{eq:I3}
\end{align}
Here again Grassmann integrals evaluate to a non-zero result only if every Grassmann variable appears exactly linearly in the integrand. Consequently, $n_k,n_l,m_i,m_j=0$, $m_k,m_l,n_i,n_j=1$, and $n_a=m_a$ if $a\neq i,j,k,l$. To reorganize the integrand into Grassmann bilinears, we use the following identity
\begin{align}
    &\Gamma_{n_in_jm_km_l}^{m_im_jn_kn_l}\alpha_i\alpha_j\alpha_k^*\alpha_l^* \overset{\rightharpoonup}{\prod_{a\neq i,j}} (\alpha_a)^{m_a}\overset{\leftharpoonup}{\prod_{b\neq k,l}} (-\alpha_b^*)^{n_b}\nonumber\\
    &=\Gamma_{n_in_jm_km_l}^{m_im_jn_kn_l} (-1)^{\varphi(i,j;\{k,l\},\boldsymbol{n}) + \varphi(k,l;\{i,j\},\boldsymbol{n})}\nonumber\\ &\quad\times \text{sgn}(i-j) \text{sgn}(k-l)(\alpha_k^*\alpha_k)(\alpha_l^*\alpha_l)\nonumber\\
    &\quad\times (\alpha_i^*\alpha_i)(\alpha_j^*\alpha_j) \overset{\rightharpoonup}{\prod_{a\neq i,j,k,l}} (\alpha_a)^{m_a}\overset{\leftharpoonup}{\prod_{b\neq i,j,k,l}} (-\alpha_b^*)^{n_b},\label{eq:I3_phases}
\end{align}
where $\text{sgn}(x)$ gives the sign of $x$ and $\varphi(i,j;\{k,l\}\boldsymbol{n})$ is defined below \Eq{eq:occ_phase}. Finally, 
\begin{align}I_3^{(i,j,k,l)}
    &= \frac{1}{4}\tilde{C}^{(4)}_{ijkl}(-1)^{\varphi(i,j;\{k,l\},\boldsymbol{n}) + \varphi(k,l;\{i,j\},\boldsymbol{n})}
    \nonumber\\
    &\quad \times\text{sgn}(i-j) \text{sgn}(k-l) \Gamma_{n_in_jm_km_l}^{m_im_jn_kn_l}\nonumber\\
    &\quad \times\prod_{p\neq i,j,k,l}\delta_{n_pm_p}[n_pg_p+(1-n_p)(1-g_p)]\label{eq:I3_fin}
\end{align}
Combining \Eqs{eq:Isum}{eq:I1_fin} yields our final result, \Eq{eq:matrix_elts} of the main text. The derivation can be generalized to include higher order correlation functions by keeping additional terms in \Eq{eq:pert_integral} and using the Grassmann identities that we presented. Such derivation is significantly simplified by a diagrammatic representation, akin to Feynmann diagram expansions, that we will illustrate below.

\subsection{Diagrammatic Rules}

\begin{figure}
    \centering
    \includegraphics[scale=.32, trim = 0 220 0 180]{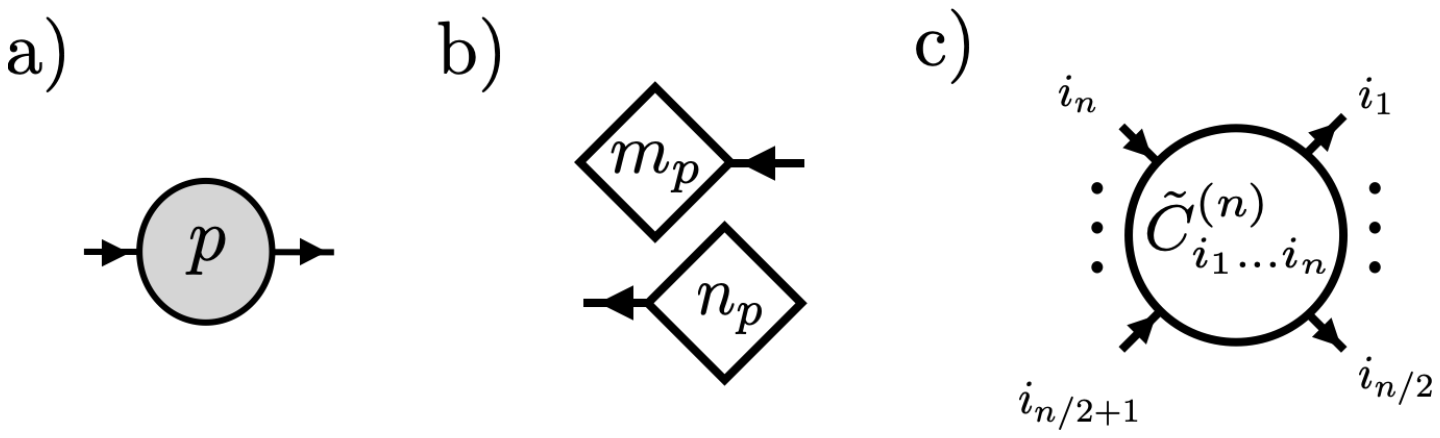}
    \caption{\textit{Elemental Diagrams.} \textbf{(a)} Propagator diagrams carrying index $p$. \textbf{(b)} Source terms, represented by diamonds carrying index $p$ with the corresponding occupations number $m_p,n_p$. \textbf{(c)} $k$-point vertex carrying indices $i_1,...,i_k$.}.
    \label{fig:elemental_diagrams}
\end{figure}

\begin{figure*}
    \centering
    \includegraphics[scale=.55, trim = 150 0 150 0]{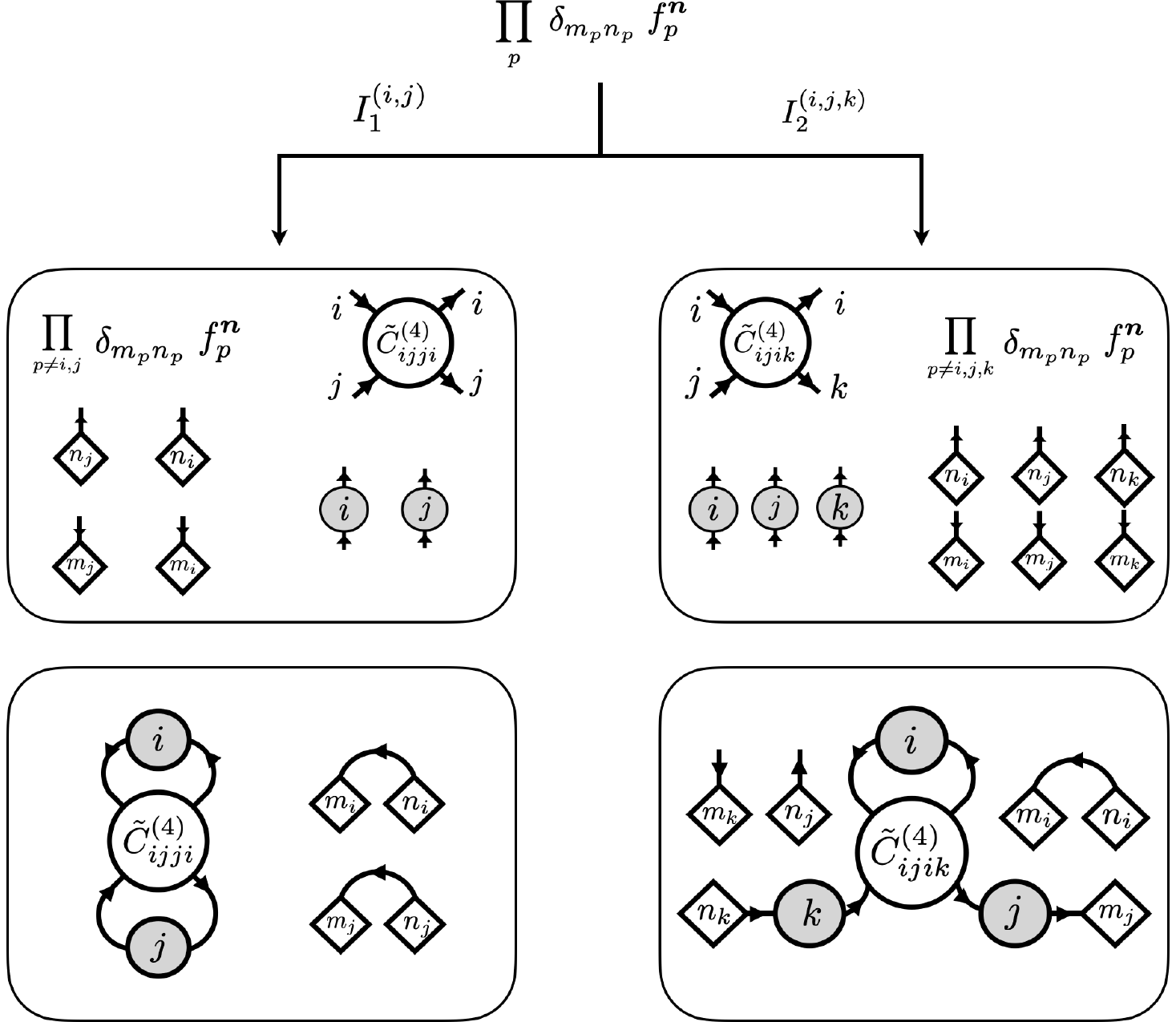}
    \caption{\textit{Construction of $I_1$ and $I_2$ Interaction Diagrams.} Starting from the Gaussian diagram as the generating functional (top), 4-point vertices are introduced. The diagram for $I_1$ is constructed on the left and the diagram for $I_2$ is constructed on the right following the rules discussed in this section.}
    \label{fig:diagram_I1I2}
\end{figure*}

\begin{figure}
    \centering
    \includegraphics[scale=.4, trim = 0 70 0 70]{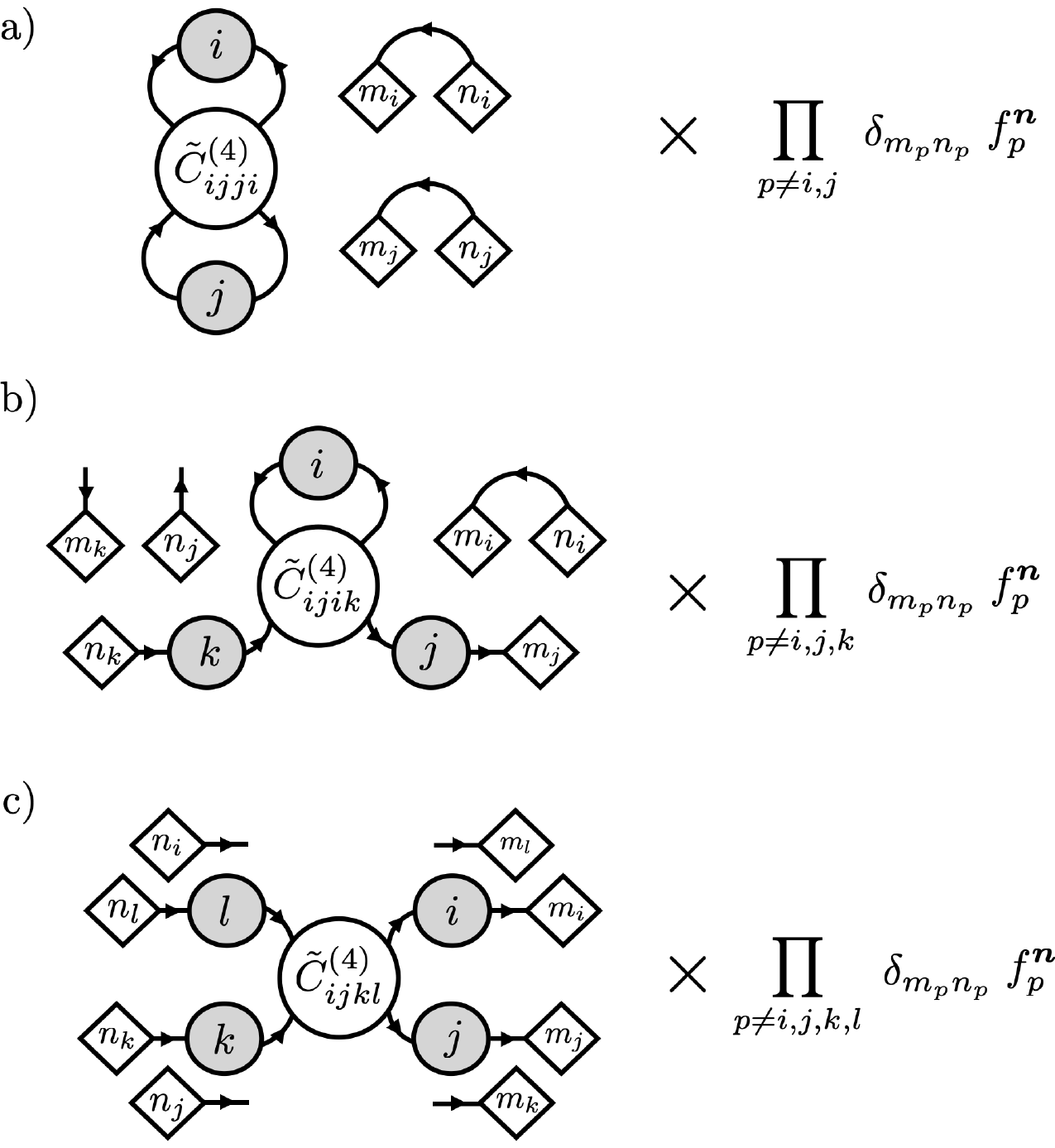}
    \caption{\textit{Diagrams contributing at $O(\tilde{C}^{(4)})$.} \textbf{(a)}  $I_1$: Propagators connect repeated indices of the vertex, all sources are paired up. \textbf{(b)} $I_2$: One propagator connects one set of repeated indices of the vertex, the other two end in sources. Some sources can be paired, others remain unconnected. \textbf{(c)} $I_3$: All propagators connect to  sources.}
    \label{fig:diagrams}
\end{figure}
\noindent
The matrix elements of the Gaussian part $\rho_{g,\boldsymbol{m}\boldsymbol{n}}$, as well as all higher-order terms in $\delta\rho_{\boldsymbol{m}\boldsymbol{n}}$ are obtained by an expansion of the fermionic Wigner functional. This involves manipulations
that contain multiple Grassmann-valued derivatives and integrals where attention must be paid to their anti-commutative nature. We now outline a diagrammatic approach that simplifies  these computations, providing intuition and allowing to extend the approach to higher-orders without loosing the rigour of the algebra that accounts for the anti-commuting nature of the variables. We will draw on analogies from Feynman diagram evaluation in Quantum Field Theory, but note that the language introduced is to build intuition only and should not be taken literally. 

The elemental components of our diagrammatic derivation is summarized in \Fig{fig:elemental_diagrams}, showing the ``propagator" in (a), ``source term'' in (b), and ``(4-point)-vertex''in (c).
We will now discuss how these elements can be used to construct the matrix elements of $\rho$. The lowest-order Gaussian part  is given by
\begin{align}
  \rho^g_{\boldsymbol{m}\boldsymbol{n}}  \equiv \prod_p \delta_{m_pn_p}f_p^{\boldsymbol{n}}\,,
\end{align}
where $f_p^{\boldsymbol{n}}\equiv n_pg_p+(1-n_p)(1-g_p)$. It will play a role akin to a generating functional for all higher order terms based on an expansion in terms of the $k$-ality of (potentially multiple) vertices. 

We refer to the subsequent terms, involving $\tilde{C}^{(4)}$, $\tilde{C}^{(6)}\dots$ and their products, as ``interaction diagrams". Their construction is illustrated pictographically in \Fig{fig:diagram_I1I2} for the leading-order case involving only $\tilde{C}^{(4)}$. Here, the Gaussian terms and $\tilde{C}^{(4)}$ serve as the starting points for these interaction diagrams. Each vertex contributes a factor ${\tilde{C}^{(k)}_{i_1,\dots,i_k}}/{(k/2)!^2}$.  To construct them, one must first distinguish between different cases of the 4-point function $\tilde{C}^{(4)}$: either two pairs of indices, one pair of indices are equal, or all indices are different. For example, the left panel of \Fig{fig:diagram_I1I2} shows the derivation of the diagram $I_1$ involving $\tilde{C}^{(4)}_{ijji}$. In this process, the ``open" indices $i \neq j$ are removed from the product over $p$ derived from the Gaussian terms. For each open index, one writes down a propagator, as well as one $\boldsymbol{n}$-type (represented by an incoming arrow on the diamond symbol) and one $\boldsymbol{m}$-type (represented by an outgoing arrow) source term for every open index. One begins by first connecting the propagators to the vertex in all possible ways so that the indices match.
Some propagators may form a closed connecting  repeated indices of $\tilde{C}^{(4)}$, in this case the diagram incurs no additional factor and the corresponding index is simply summed over.
Otherwise, a propagator's end point must be connected to an $\boldsymbol{n}$- or $\boldsymbol{m}$-type source. Finally, one directly connects  $\boldsymbol{m}$- and $\boldsymbol{n}$-type source terms if their indices match, in this case they contribute a factor of $(-1)^{n_a}\delta_{m_an_a}$. Some source terms may be left unmatched,  see for instance the bottom right panel of \Fig{fig:diagram_I1I2} where $m_k$ and $n_j$ with $k\neq j$ are unmatched. In this case, they result in a factor of $\Gamma^{n_km_j}_{m_kn_j}$. The left panel of \Fig{fig:diagram_I1I2} illustrates the unique way
that $\tilde{C}^{(4)}_{ijji}$ can be connected, while the right panel of the same figure depicts the diagram for the term $I_2$ which involves the ``pair-density" correlation function $\tilde{C}^{(4)}_{ijik}$; the Gaussian contribution is not explicitly shown.

Additionally, because of the anti-commutative nature, we must account for phases that originate from un-paired sources (with open arrows). We denote the ordered set of in-going indices as $J_{\rm in}$, the ordered set of out-going indices as $J_{\rm out}$, and $J=J_{\rm out}\cup J_{\rm in}$ as their union, where the ordering is the same as how the indices appear in the corresponding $k$-point vertex $\tilde{C}^{(k)}_{i_1,...,i_k}$. For example, for the diagram depicting $I_2$, shown in the right panel of \Fig{fig:diagram_I1I2}, $J_{\rm in}=\{ k \} $ and $J_{\rm out}=\{ j \} $ while for $I_3$ $J_{\rm in}=\{ k,l \} $ and $J_{\rm out}=\{ i,j \} $. We then define $\tilde{\varphi}=\sum_{f,g\in J}\varphi(f,g;J/\{f,g\},\boldsymbol{n})$ which sums the occupations of $\boldsymbol{n}$ between all pairs $(f,g)$ of external indices regardless of whether they are in-going or out-going, ignoring the occupations of the external indices not in the pair denoted by $J/\{f,g\}$. The diagram is multiplied by a phase factor $\Phi$ which is
\begin{align}
    \Phi = (-1)^{\tilde{\varphi}} &\prod_{a<b<\#J_{\rm in}} \text{sgn}\left(J_{\rm in,a}-J_{\rm in,b}\right)\nonumber\\
    & \times \prod_{a<b<\#J_{\rm out}}\text{sgn}\left(J_{\rm out,a}-J_{\rm out,b}\right)
\end{align}
where $J_{\rm in,a}$ and $J_{\rm out,a}$ denote the $a^{th}$-element of $J_{\rm in}$ and $J_{\rm out}$ respectively, $\#$ is the cardinality of a set, and $\text{sgn}(x)$ gives the sign of $x$.
 
\begin{figure}[t]
    \centering
    \includegraphics[scale=.4]{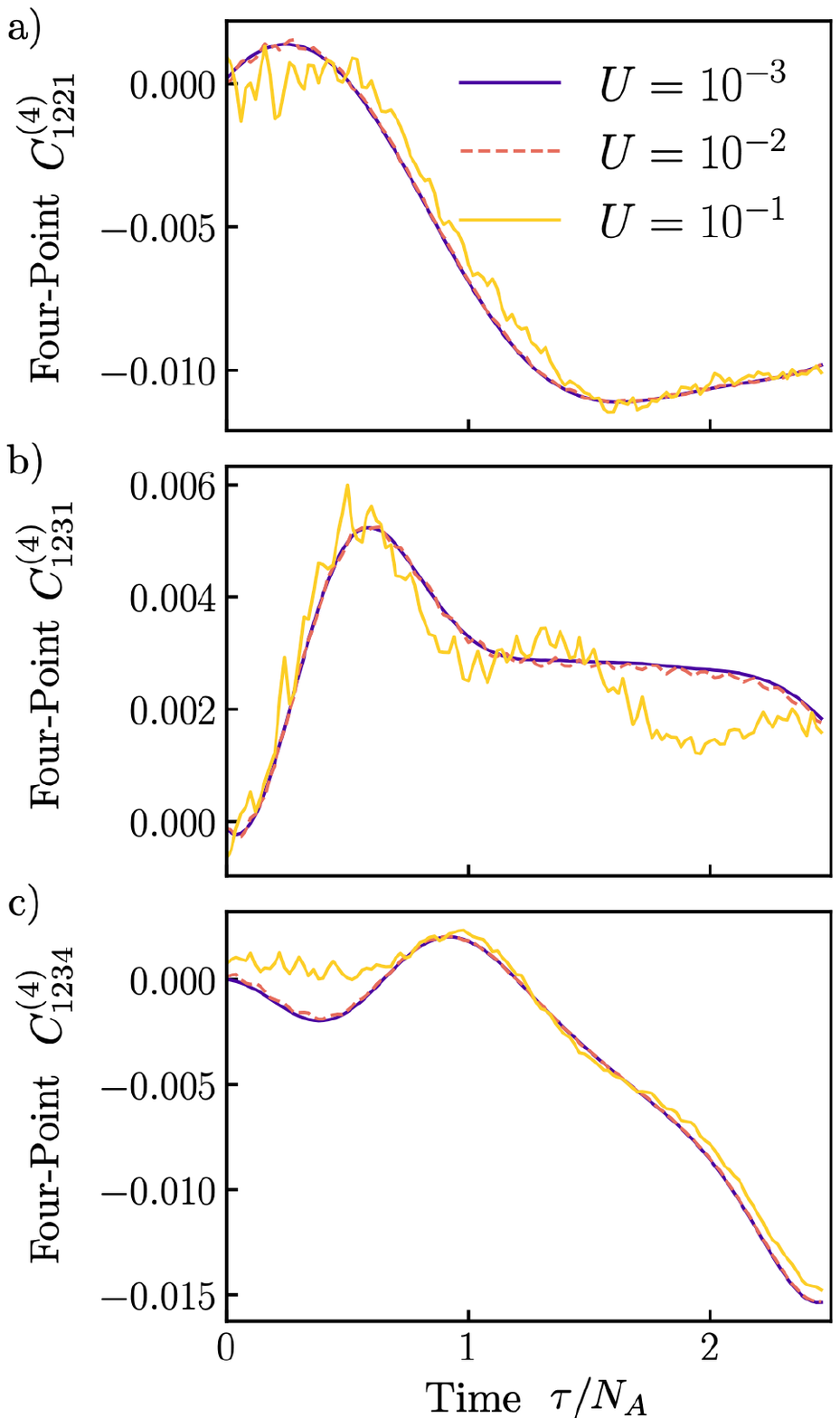}
    \caption{\textit{Self-similarity of non-Gaussian correlations.} \textbf{(a)} Example of a  ``density-density" correlation, $\tilde{C}^{(4)}_{ijji}$ with $i \neq j$, for the same interaction strengths as shown in \Fig{fig:fig2} of the main text. For $U=10^{-3},10^{-2}$, the correlations collapse onto the same curve when rescaled in by the interaction strength. As $U\sim J,J'$ oscillations begin to appear on top of this curve that breaks this behavior. \textbf{(b)} Example of a ``pair-density" correlation function $\tilde{C}^{(4)}_{ijki}$ with $i \neq j \neq k$. \textbf{(c)} Example of a  ``pair-pair" correlation function ($i \neq j \neq k \neq l$).}
    \label{fig:fig7}
\end{figure}

\subsection{Properties of the Parameterization}
\noindent
We now discuss the parameterization's properties, including its spectrum and matrix structure. The non-Gaussian contribution derived in this manuscript  is Hermitian $\delta\rho=\delta\rho^{\dagger}$ and has zero trace $\text{Tr}\left[\delta\rho\right]=0$. However, it does not constitute a bona fide quantum state, as $\rho + \delta\rho $ is not always a positive operator if only $\tilde{C}^{(4)}$ is included.   This is a consequence from feeding only partial information into the parameterization: Generically, if $\tilde{C}^{(4)}$ increases, so will higher-order correlations, $\tilde{C}^{(6)}$, $\tilde{C}^{(8)}\dots$. If they are not included, the magnitudes of negative eigenvalues increases with   $\tilde{C}^{(4)}$, indicating that the approximation being used becomes invalid. Including them will lead to a positive semi-definite matrix.  In practice this is a minimal numerical effect when used within the applicability of the ansatz: Even for small but non-trivial $\tilde{C}^{(4)}$, $\delta\rho$ there may be few negative eigenvalues, although with extremely small magnitudes. Nevertheless, this can cause problems when computing the entanglement Hamiltonian numerically by taking the matrix-logarithm of a state. We resolve this issue and define a proper state by diagonalizing the matrix and projecting the Schmidt spectrum onto the closest classical probability distribution under the 2-norm using the approach of~\cite{acharya2021informationally}.

The non-Gaussian part $ \delta\rho $ has diagonal and off-diagonal components. The diagonal part includes contributions from connected 4-point functions corresponding to ``density-density" correlations, specifically $\tilde{C}^{(4)}_{ijji}$ with $i \neq j$. On the other hand, ``pair-density" correlations, such as $\tilde{C}^{(4)}_{ijki}$ with $i \neq j \neq k$, contribute to off-diagonal elements, where the occupations $\boldsymbol{m}$ and $\boldsymbol{n}$, when written as a bitstring, differ by a Hamming distance of $d(\boldsymbol{m},\boldsymbol{n})=2$. Additionally, ``pair-pair" correlations, $\tilde{C}^{(4)}_{ijkl}$ with $i \neq j \neq k \neq l$, connect quasi-particle states with a Hamming distance of $d(\boldsymbol{m},\boldsymbol{n})=4$. In our numerical simulations we observed that including only the diagonal two-body 'density-density' correlations is insufficient to produce level repulsion when the underlying state behaves chaotically. We attribute this behaviour to uncorrelated neighboring eigenvalues (as in a Poissonian distribution), which is preserved by the diagonal two-body correlations. Thus the ability to measure off-diagonal correlations, not just simple non-Gaussian density correlations, is important to diagnose the phenomena we study in the main text.

\section{Early Time Growth of Non-Gaussianity}\label{app:early_time}

In this section of the Supplemental Material, we provide additional quantitative details on the early-time dynamics of the quantum quench studied in the main text. Specifically, we examine the dynamics of our measure of non-Gaussianity by comparing the deviation between the exact state, $\theta(\rho_A)$, and our parameterization, $\theta(\tilde{\rho})$, from a Gaussian state. As shown in \Fig{fig:fig2}(a) and (b), we observed that for small couplings $U \ll J, J'$, $\theta$ grows linearly and exhibits self-similarity during the early times.
As noted in the main text, both $\theta(\rho_A)$ and $\theta(\tilde{\rho})$ behave similarly in this regime, so our discussion focuses on $\theta(\rho_A)$. Specifically, without loss of generality, we consider the dynamics of the full $\rho$ instead of the reduced $\rho_A$ because we have checked that $\theta(\rho)$ exhibits the same linear growth as $\theta(\rho_A)$.

The system is composed of $N$ sites with interactions described by Hamiltonian $H=H_0+H_I$, where $H_0=J\sum_{l,s}c^{\dagger}_{l+1s}c_{ls} + J'\sum_{l,s}c^{\dagger}_{l+2s}c_{ls} + h.c.$ and $H_I=U\sum_ln_{l\uparrow}n_{l\downarrow}$. The full system is translationally invariant with periodic boundary conditions, and so we define momentum-space mode operators 
\begin{align}
    b^{\dagger}_{k\sigma}=\frac{1}{\sqrt{N}}\sum_{l}e^{ikl}c^{\dagger}_{l\sigma}\,.
\end{align}
Because $\rho$ is pure $\text{Tr}[\rho^2]=1$, allowing us to write the fidelity t as $\mathcal{F}(\rho|\rho_g)=\text{Tr}[\rho\rho_g]/\text{max}\{\text{Tr}[\rho_g^2],\text{Tr}[\rho^2]\}=\text{Tr}[\rho\rho_g]$, where $\rho_g$ is the Gaussian state with the same two-point correlations as the exact state. In this study, the quench begins from the intial plane wave state, 
\begin{align}\label{eq:initialstatedef}
    \rho(0)=\ket{\boldsymbol{n}_0}\bra{\boldsymbol{n}_0}\,,\qquad \ket{\boldsymbol{n}_0}=\prod_{k}(b^{\dagger}_{k\sigma})^{(\boldsymbol{n}_0)_{k\sigma}}\ket{0}
\end{align}   
where $\boldsymbol{n}_0$ is the initial occupation of momentum modes written as a bitstring and $| 0\rangle$ is the Fock vacuum. The fidelity, as a function of evolution time, is
\begin{align}
    \mathcal{F} &= \braket{\boldsymbol{n}_0|e^{i(H_0+H_I)t}\rho_g(t)e^{-i(H_0+H_I)t}|\boldsymbol{n}_0}\nonumber\\
    &= \braket{\boldsymbol{n}_0|e^{iH_It}\rho_g(t)e^{-iH_It}|\boldsymbol{n}_0} + \mathcal{O}(JUt^2)\label{eq:earlytime_fid}\,.
\end{align}
In the second line, we separated the time evolution operator into its free and interacting components, using the first-order Trotter-Suzuki formula, and  we used the fact that $\ket{\boldsymbol{n}_0}$ is an eigenstate of the free Hamiltonian $H_0$. Likewise, the Gaussian state may be written as $\rho_g=[\prod_{k,\sigma}g_{k\sigma}]\,\text{exp}\{\sum_{k,\sigma}\log[g_{k\sigma}(1-g_{k\sigma})^{-1}]b_{k\sigma}^{\dagger}b_{k\sigma}\}$ where the indices $k\in\{-\pi,...,\pi-2\pi/N\},\sigma\in\{\uparrow,\downarrow\}$. Here, the two-point function $\braket{b_{k\sigma}^{\dagger}b_{k'\sigma'}}=g_{k\sigma}\delta_{kk'}\delta_{\sigma\sigma'}$ is diagonal in the plane wave basis for translationally invariant systems. The early time behavior of the $g_{k\sigma}$ can be found by solving the Heisenberg equations of motion  $\partial_t (b_{k\sigma}^{\dagger}b_{k\sigma})=i[H_0+H_I,b_{k\sigma}^{\dagger}b_{k\sigma}]$. In the plane wave basis $H_0=\sum_{k\sigma}\varepsilon_{k}b_{k\sigma}^{\dagger}b_{k\sigma}$ and \[H_I=\frac{U}{N}\sum_{k_1,k_2,k_3,k_4}\delta(k_1+k_2-k_3-k_4)b_{k_1,\uparrow}^{\dagger}b_{k_2,\downarrow}^{\dagger}b_{k_1,\downarrow}b_{k_3,\uparrow}\] with $\varepsilon_{k}=2[Jcos(k)+J'cos(2k)]$, and so integrating the equations of motion gives
\begin{align}
    g_{k\sigma}(t) = g_{k\sigma}(0) + &\frac{iU}{N}\int_0^tds \sum_{k_1,k_2,k_3,k_4}\delta(k_1+k_2-k_3-k_4)\nonumber\\
    &\times \braket{[b_{k_1,\uparrow}^{\dagger}b_{k_2,\downarrow}^{\dagger}b_{k_1,\downarrow}b_{k_3,\uparrow},b_{k\sigma}^{\dagger}b_{k\sigma}]}(s)\nonumber\\
    \approx (\boldsymbol{n}_0)_{k\sigma}\;+&\;\frac{iUt}{N}G_{k\sigma}
\end{align}
where we have used the fact the initial state is $\ket{\boldsymbol{n}_0}$ and \begin{align*}G_{k\sigma}&=\sum_{k_1,k_2,k_3,k_4}\delta(k_1+k_2-k_3-k_4)\\ &\qquad
\braket{[b_{k_1,\uparrow}^{\dagger}b_{k_2,\downarrow}^{\dagger}b_{k_1,\downarrow}b_{k_3,\uparrow},b_{k\sigma}^{\dagger}b_{k\sigma}]}(0) \;.\end{align*} In defining $G_{k\sigma}$ we have treated the commutator term as stationary on the timescales considered, which is an approximation that is used in the literature \cite{stark2013kinetic}. With knowledge of $g_{k\sigma}$, at early times we may express the Gaussian state as $\rho_g(t)=B(t)\ket{\boldsymbol{n}_0}\bra{\boldsymbol{n}_0}B^{\dagger}(t)$ with $B(t)=\text{exp}\{\frac{iUt}{N}\sum_{k,\sigma}G_{k\sigma}/[(\boldsymbol{n}_0)_{k\sigma}(1-(\boldsymbol{n}_0)_{k\sigma})]b^{\dagger}_{k\sigma}b_{k\sigma}\}$, which comes from expanding $\log[g_{k\sigma}(1-g_{k\sigma})^{-1}]$ for small $Ut$. Thus \Eq{eq:earlytime_fid} becomes
\begin{align}
    \mathcal{F} &= \bra{\boldsymbol{n}_0}e^{iH_It}B(t)\ket{\boldsymbol{n}_0}\bra{\boldsymbol{n}_0}B^{\dagger}(t)e^{-iH_It}\ket{\boldsymbol{n}_0} + \mathcal{O}(JUt^2)\nonumber\\
    &= \bra{\boldsymbol{n}_0}(1+iH_It+\frac{iUt}{N}\sum_{k\sigma}G_{k\sigma}b_{k\sigma}^{\dagger}b_{k\sigma})\ket{\boldsymbol{n}_0}\nonumber\\
    &\times\bra{\boldsymbol{n}_0}(1-iH_It-\frac{iUt}{N}\sum_{k\sigma}G_{k\sigma}b_{k\sigma}^{\dagger}b_{k\sigma})\ket{\boldsymbol{n}_0} + \mathcal{O}(\tau^2)\nonumber\\
    &= 1 + \mathcal{O}(\tau^2)\,,
\end{align}
where on the second line we have expanded the exponents for small $Ut$. 

In conclusion, we have shown that the terms linear in $\tau=Ut$ cancel, and so the fidelity $\mathcal{F}$ will decrease from $1$ at most quadratically in $\tau$. Because we defined the distance $\theta(\rho)=\text{arccos}\sqrt{\mathcal{F}(\rho|\rho_g)}$, we expect that $\theta(\rho)$ grows linearly at early times.

We attribute the observed self-similarity to the system's  two-body (``four-point") correlations, although the exact mechanism remains unknown to us at this point and deserves further study. To illustrate this effect, \Fig{fig:fig7} shows selected four-point functions used in the calculation of $\theta(\tilde{\rho})$ as functions of $\tau$ for the same values of $U$ as in Figure \Fig{fig:fig2} (a) and (b) in the main text. For $U \ll J, J'$, the correlations collapse onto the same curve. However, for $U \sim J, J'$, oscillations on this curve begin to grow until all resemblance to the small $U$ correlations is lost.

\section{Discussion of Symmetries and choice of the Initial State }\label{app:init_state}
We now discuss the  symmetries, relevant for this study, of the Fermi-Hubbard model with a next-nearest-neighbor hopping \cite{essler2005one}. One $U(1)$ symmetry  is  local particle number conservation,
\begin{equation}\label{eq:spin_operator}
    N=\sum_ln_{l,\uparrow}+n_{l\downarrow}\,.
\end{equation}
For a subsystem of size $N_A$, there are $2N_A+1$ particle number sectors with associated quantum numbers $n\in\{0,...,2N_A\}$. Additionally, the model has an $SU(2)$ symmetry, generated by the spin operators
\begin{equation}\label{eq:spin_operator}
    S^i=\frac{1}{2}\sum_l\sum_{r,r'}c^{\dagger}_{l,r}\left(\sigma^i\right)_{rr'}c_{l,r'}
\end{equation}
where $l\in\{0,...,N_A-1\}$, $r,r'\in\{\uparrow,\downarrow\}$ and $\vec{\sigma}=\{\sigma^x,\sigma^y,\sigma^z\}$ is a vector of Pauli matrices. This symmetry corresponds to global angular momentum conservation. The quantum numbers associated with this symmetry are the eigenvalues of the total magnetization operator $\hat{S}^z=\frac{1}{2}\sum_ln_{l\uparrow}-n_{l\downarrow}$ and the Casimir operator $S^2=(S^x)^2+(S^y)^2+(S^z)^2$. There are $2N_A+1$ quantum number sectors associated with $S^z$, denoted by $m\in\{-N_A,...,N_A\}$ and $N_A+1$ quantum numbers associated to $S^2$, denoted by $s\in\{0,...,N_A(N_A+1)\}$. The three operators $N, S^z$ and $S^2$ form the symmetry operators whose eigensectors are important for analyzing the dynamics of this model in this paper. In practice, one must project the state $\rho_A(t)$ into these symmetry sectors when analysing the level distribution of the entanglement Hamiltonian, because eigenvalues from different sectors remain trivially uncorrelated when also the initial state respect these symmetries. 

There is also a particle-hole like symmetry that exists at half-filling, i.e. $n=N_A$ (see~\cite{de2022level}), however its details are largely irrelevant for our purposes. Additionally, modifications of the model may feature additional symmetries, such as e.g., non-Abelian $\eta$-pairing symmetry~\cite{de2022level}, however they are either not relevant for us or explicitly broken. Space-time symmetries  are broken explicitly when one restricts the state to a subsystem.

In this study, we focus on
initial momentum-space Fock states given by \Eq{eq:initialstatedef}.
We now discuss how to choose such an initial state for the quench experiment that allows measurement of the build-up of level repulsion.  Because studying level repulsion requires block-diagonalizing the state into symmetry sectors, and because the model has so many symmetries reflected in the measured correlation functions, one option is to resolve all of them when analyzing $\rho_A$. Another is to break the symmetry explicitly by the initial condition. Opting for a combination of the two, we will break  the $S^2$ and particle-hole symmetries of the model by choice of a suitable initial state.
\begin{figure}[t]
    \centering
   \includegraphics[width=0.45\textwidth, trim = 0 70 0 70]{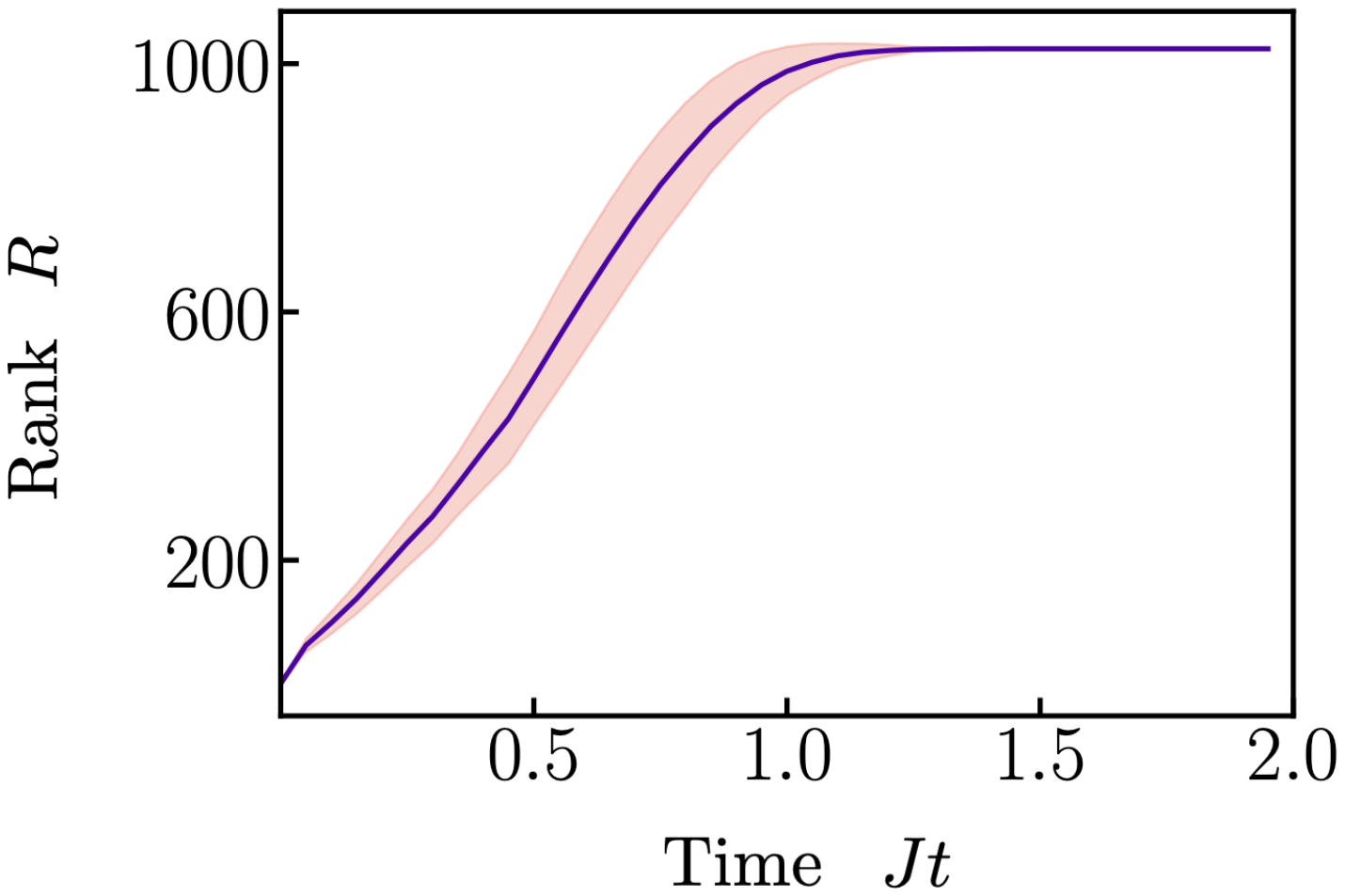}
    \caption{\textit{Rank of the state.} Growth of rank for a system of $N=10$ during the free evolution averaged over 30 random initial states. The pink region shows the range of growth rates over the averaged states.}
    \label{fig:rank_plots}
\end{figure}
The particle-hole is broken by choosing an initial state that is not at half-filling. The $S^2$ symmetry may be broken by choosing a state that is not an eigenstate of the Casimir, i.e. $\mathcal{O}_0=[\rho_0,S^2]\neq 0$. The algorithm for finding such initial state is as follows:
\begin{enumerate}
    \item Choose a global number sector $n$ that is away from half-filling (i.e. $n\neq N_A$).

    \item Randomly generate a trial product state in the Fock basis, \Eq{eq:initialstatedef}, that satisfies $\sum_{k\sigma}(\boldsymbol{n}_0)_{k\sigma}=n$

    \item Check $\text{Tr}[\mathcal{O}_0^{\dagger}\mathcal{O}_0]\neq 0$. If this condition is satisfied, accept this initial state and proceed with time evolution. If it is not, return to step 2.
\end{enumerate}

\begin{figure}[t]
    \centering
    \includegraphics[width=0.9\columnwidth]{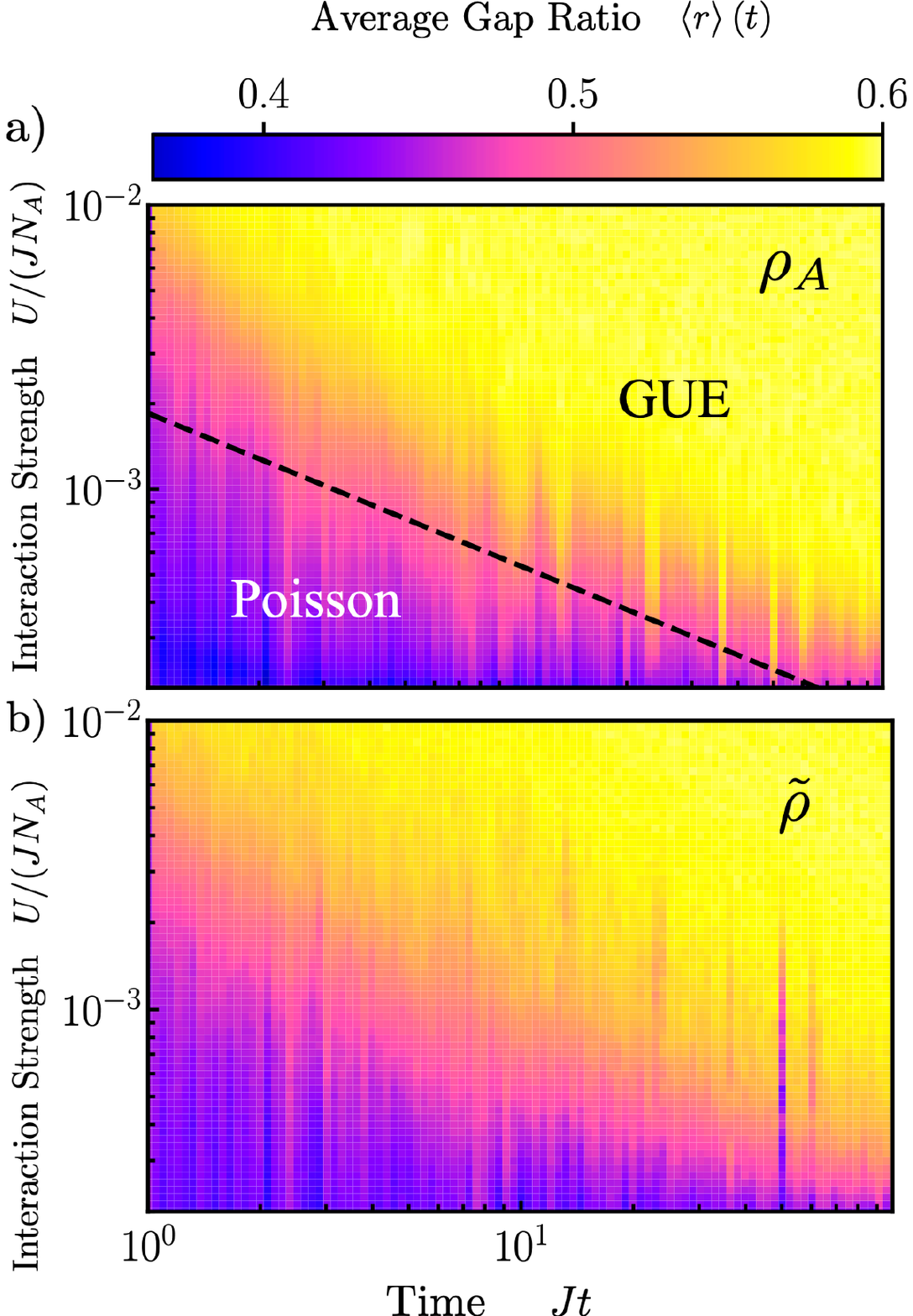}
    \caption{\textit{Position space initial state.} \textbf{(a)} The average gap ratio of the exact state with a postion space initial state. \textbf{(b)} The average gap ratio $\braket{r}(t)$ predicted by the parameterization.}
    \label{fig:r_scan_loc}
\end{figure}
An alternative strategy is to start from initial \textit{position-space} Fock states
\begin{align}
     \ket{\tilde{\boldsymbol{n}}_0}=\prod_{k}(c^{\dagger}_{k\sigma})^{(\tilde{\boldsymbol{n}}_0)_{k\sigma}}\ket{0}
\end{align}
which are un-entangled product states, making their experimental preparation simpler. Symmetries can be adjusted in the exact same way as for momentum-space states. Since such state is a product state and therefore a very specific type of Gaussian state, it has zero entanglement and an initially trivial entanglement Hamiltonian with rank $\mathcal{R}_A = 1$. Consequently, our primary observable, the entanglement gap ratio, is initially undefined. To enable comparison with a generic  Gaussian state, we first evolve the state with the free Hamiltonian $H_0$ for a duration $t_0$ before turning-on interactions, i.e.
\begin{align}
    \ket{\psi}(t) &= 
    \begin{cases}
        e^{-itH_0}\ket{\tilde{\boldsymbol{n}}_0} & \text{for } t<t_{0}\\
        e^{-i(t-t_{0})(H_0+H_I)}e^{-it_{0}H_0}\ket{\tilde{\boldsymbol{n}}_0} & \text{for } t\geq t_{0}
    \end{cases}
\end{align}
During the onset period, we monitor the effective rank $\mathcal{R}_A\equiv \lim_{\alpha\rightarrow 0}\exp\{\frac{1}{1-\alpha}\log(\text{Tr}[\rho^\alpha]) \}$ 
of the bipartite reduced density matrix. It increases until it reaches full rank while remaining a Gaussian state, see~\Fig{fig:rank_plots}. We ensure that $t_0$ is always chosen large enough to achieve full rank before turning on interactions. In experiment, the turning on of interactions could be realized by keeping the two fermion species in separate lattice first before bringing them together using a reconfigurable array of optical tweezers or moving a spin-selective optical lattice.

After the interactions are turned on, our analysis is identical to that in the main text. \Fig{fig:r_scan_loc} shows the average gap ratio over a range of interaction strengths for both $\rho_A$ and $\tilde{\rho}$. Even for large $U$ the parameterization  captures the level statistics of the exact state, and accurately maps the cross over between uncorrelated levels (Poisson) and level-repulsion (GUE).

\section{Cost, efficiency, and noise}\label{app:cost}
The method proposed here enables the reconstruction of an approximate quantum state using lower-order correlation functions. Applications—such as extracting the level distribution of an associated entanglement Hamiltonian—incur costs in both the sampling complexity of measuring the correlation functions and in the classical postprocessing required. These procedures are also sensitive to noise. In the following, we address these challenges and note that strict complexity-theoretic performance guarantees remain an open question warranting further investigation.

We begin by addressing the cost associated with classical postprocessing—specifically, whether our application requires storing the (exponentially large) reduced density matrix on a classical computer in order to compute its spectrum. In principle, it is not necessary to store the full matrix to access spectral information. Instead, one can apply generalized degenerate perturbation theory to an order consistent with the included correlation functions. To illustrate this approach, we combine the Gaussian part with the diagonal contribution of $\delta \rho$ into a single diagonal matrix,
\begin{align}
D_{\mathbf{m}\mathbf{n}} = \delta_{\mathbf{m}\mathbf{n}} (\rho_{g,\mathbf{n}\mathbf{n}} + \delta \rho_{\mathbf{n}\mathbf{n}}) \;.
\end{align}

%
%We now give a cost analysis of our method applied to the target application of reconstructing the spectral statistics of the EH, first addressing the cost of classically reconstructing the reduced state. In our method, the exponential cost of having to store the entire density matrix in memory can in principle be circumvented by targeting specific levels relevant for a desired observable with generalized degenerate perturbation theory to an order consistent with the correlations being included. Specifically, we collect both the Gaussian part together with the diagonal contribution of $\delta \rho$ in the diagonal matrix}
%\begin{align}
%D_{\mathbf{m}\mathbf{n}} = \delta_{\mathbf{m}\mathbf{n}} %(\rho_{g,\mathbf{n}\mathbf{n}} + \delta \rho_{\mathbf{n}\mathbf{n}}) \;.
%\end{align}
We work in the occupation basis where $\rho_g$ is diagonal and $\delta \rho_{\mathbf{n}\mathbf{n}}$ are the corresponding contributions with $d_H(\mathbf{n},\mathbf{n}) = 0$ from Eq.~(5) of the main text. The remaining contributions with $d_H(\mathbf{m},\mathbf{n}) = 2$ or $4$ form the perturbation $V_{\mathbf{m} \mathbf{n}}$. 
We can thus calculate the perturbed eigenvalues $D'_{\mathbf{n}\mathbf{n}}$ of $\rho = D + V$ as
\begin{align}\label{eqn:pert_eqn}
D'_{\mathbf{n}\mathbf{n}} = D_{\mathbf{n}\mathbf{n}} + \sum_{\mathbf{m}\neq \mathbf{n}} \frac{V_{\mathbf{n}\mathbf{m}} V_{\mathbf{m}\mathbf{n}}}{D_{\mathbf{n}}- D_{\mathbf{m}}} \;.
\end{align}
The sum over all occupation numbers $\mathbf{m}$ is \emph{not} exponentially large in the subsytem size because for any given $\mathbf{n}$ only contributions with $d_H(\mathbf{m},\mathbf{n}) = 2$ or $4$ from the four-point functions arise, which reduces the amount of terms to be computed to a number polynomial in subsystem size, specifically $O(N_A^4)$. This shows that one can have access to the entanglement spectrum without classically reconstructing the reduced state as a matrix.

Proceeding this way allows for the estimation of certain eigenvalues-for example the lowest lying levels- at a given order, it is insufficient to capture level repulsion at low orders in perturbation theory, see e.g.~\cite{szasz2021weak}. The entire reduced state must be explicitly constructed from our parameterization and diagonalized which incurs a cost exponential in subsystem size, though importantly does not scale with the size of the full system. This holds despite the fact that two- and four-point functions carry sufficient information to discern level repulsion as we show in the main text. Fortunately, since level repulsion is a universal quantity, it is sufficient to choose a subsystem large enough to yield meaningful level statistics, but still small enough to be classically representable.

\begin{figure}
    \vspace*{1em}
    \centering
\includegraphics[scale=.4,width=\linewidth]{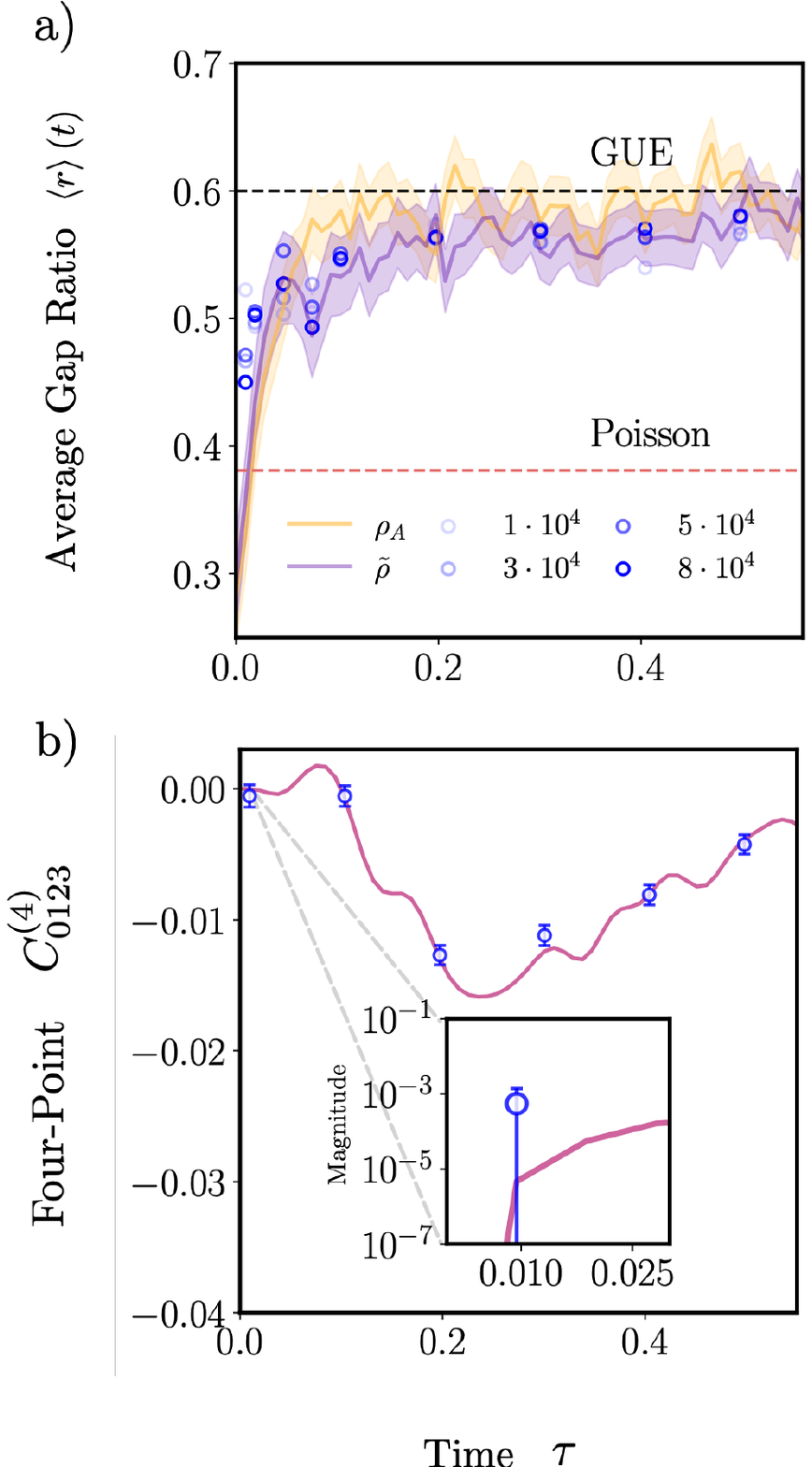}
    \caption{\textit{Shot noise analysis.} \textbf{(a)} The build-up of level repulsion for $N=8,N_A=4$ averaged across 4 initial states. The gold line represents the exact result, the purple line represents the ansatz with exact four-point functions, and the blue dots denote the ansatz measured with finite shots; the shading indicates the number of samples used. \textbf{(b)} An example of four-point function (points) compared to the exact four-point function (continuous curve). The error bars are given by the standard deviation from the mean and the number of shots are $N_S=5\cdot10^4$}
    \label{fig:shots}
\end{figure}

Second, we now address the effect of shot noise on reconstructing the spectral statistics.  In Fig.~\ref{fig:shots} (a) we emulated the measurement protocol discussed in the main text for $4$ initial states at times $\tau=0.01,0.02,0.05,0.08,0.1,0.2,0.3,0.4,0.5$ for four runs with $N_S=10^{4},3\cdot10^4,5\cdot10^4,8\cdot10^4$, where $N_S$ is the number of shots for each basis in which correlation functions are extracted, for each run. After times $\tau\geq.05$, we observe that the gap ratio from the estimated four point functions is able to closely track the gap ratio from the exact four-point functions. However, for the first few time steps the convergence to the exact prediction is observed to be much slower, behavior that can be attributed to the typical size of non-Gaussian correlations at very early times.

Fig.~\ref{fig:shots} (b) depicts the real part of an example four-point function with the associated finite-shot estimate at $\tau = 0.01,0.1,0.2,0.3,0.4,0.5$, which for $\tau=0.01$ is of the order $\mathcal{O}(10^{-6})$ while the shot-noise estimated value is $\mathcal{O}(10^{-3})$, both smaller than the two-point functions which are $\mathcal{O}(10^{-1})$. % Experimental imperfections and shot noise
%generally overestimate high-order correlation functions, for instance strength of the four-point function compared to Gaussian correlations, and thus lead to overestimation of level repulsion. 
At later times the four-point function has grown sufficiently such that it can be accurately measured while still being small enough for the parameterization to remain valid. %Thus, it is seen that even though very early time regimes cannot be accessed, one is still able to probe the gap ratio at times preceding the complete transition to GUE, allowing insight into the timescales relevant for quantum thermalization.}

In general, experimental imperfections and shot noise tend to overestimate higher-order correlation functions relative to those of Gaussian states  which are structurally special. This leads to an overestimation of level repulsion. Taken together, these observations highlight that estimating entanglement-related properties of near-Gaussian states is a subtle and nuanced task. While initial complexity-theoretic results provide valuable insights~\cite{gu2024simulating,bittel2024optimal}, this merits further investigation.

\end{document}